\documentclass[12pt]{article}
\pdfoutput=1
\usepackage{soul}
\usepackage{setspace}
\usepackage{slashed}
\usepackage{graphicx,url}
\usepackage{booktabs}
\usepackage{bbm}
\usepackage{braket}
\usepackage{amsmath,amssymb,cancel,stmaryrd}
\usepackage[usenames,dvipsnames]{xcolor}
\usepackage{color} 
 
\usepackage{epsfig}
\usepackage{array,multirow}
\usepackage{jheppub}
\def\half{{1 \over 2}}

\def\lb{\bar{\lambda}}  
 
\def\mb{\bar{\mu}}

\def\Or[#1]{{\text{O}}\left({#1}\right)}
\def\dotl[#1,#2]{\left\langle #1, #2 \right\rangle}
\def\dotlb[#1,#2]{[ #1, #2 ]}
\def\dotp[#1,#2]{(#1) \cdot (#2)}
\def\aff[#1,#2]{\hat{#1}(#2)}
\def\n4sym{{\cal N}=4 SYM}
\def\>{\rangle}
\def\<{\langle}
\def\weight[#1,#2,#3]{\{(#1),#2,#3\}}
\def\ads[#1]{$\text{AdS}_{#1}$}

\newcommand{\ba}{\begin{eqnarray}}
\newcommand{\ea}{\end{eqnarray}}
\newcommand{\be}{\begin{eqnarray}}
\newcommand{\ee}{\end{eqnarray}}
\newcommand{\bq}{\begin{equation}}
\newcommand{\eq}{\end{equation}}
\newcommand{\bal}{\begin{aligned}}
\newcommand{\eal}{\end{aligned}}
\newcommand{\benn}{\begin{equation*}}
\newcommand{\eenn}{\end{equation*}}
\newcommand{\bi}{\begin{itemize}}  
\newcommand{\ei}{\end{itemize}}

\newcommand{\Bcal}{{\mathcal B}}

\newcommand{\Ocal}{{\mathcal O}}

\newcommand{\CL}{{\cal L}}

\newcommand{\CO}{{\cal O}}

\newcommand{\nn}{\nonumber}

\newcommand\oo\infty
\newcommand\s\sigma
\newcommand\de\delta
\newcommand\De\Delta
\newcommand{\p}{\partial}
\newcommand\f\phi
\newcommand\g\gamma
\newcommand\x\times

\newcommand{\ra}{\rightarrow}

\newcommand{\fr}{\frac}

\newcommand{\eff}{\textrm{eff}}
\newcommand{\gap}{\textrm{gap}}
\newcommand{\ET}{\textrm{ET}}
\newcommand{\LC}{\textrm{LC}}

\newcommand{\Dmax}{\De_{\max}}
\newcommand\lrpar{\raise .8ex\hbox{$^\leftrightarrow$} \hspace{-9pt}
\partial}

\newcommand\G{\Gamma}

\newcommand{\tB}{\widetilde{\Bcal}}

\newcommand{\norder}[1]{%
  {:\mathrel{\mspace{1mu}#1\mspace{1mu}}:}%
}

\makeatletter
\def\@fpheader{\vspace{-.1cm}}
\makeatother

\preprint{CERN-TH-2018-277}

\title{
Nonperturbative Matching Between Equal-Time and Lightcone Quantization
}

\author[a]{A.~Liam Fitzpatrick,}
\author[a]{Emanuel Katz,}
\author[b]{Matthew T. Walters}

\affiliation[a]{Department of Physics, Boston University, \\
Commonwealth Avenue, Boston, MA 02215, USA}

\affiliation[b]{CERN, Theoretical Physics Department, \\
1211 Geneva 23, Switzerland}

\abstract{We investigate the nonperturbative relation between lightcone (LC) and standard equal-time (ET) quantization in the context of $\lambda \phi^4$ theory in $d=2$. We discuss the perturbative matching between bare parameters and the failure of its naive nonperturbative extension.  We argue that they are nevertheless the same theory nonperturbatively, and that furthermore the nonperturbative map between bare parameters can be extracted from ET perturbation theory via Borel resummation of the mass gap.    
We test this map by using it to compare physical quantities computed using numerical Hamiltonian truncation methods in ET and LC.  
}

\arxivnumber{}

\begin{document}

\maketitle  



\section{Introduction and Summary}

Quantization on surfaces of constant lightcone (LC) time $x^+ \equiv \fr{1}{\sqrt{2}}(t+x)$ leads to a number of simplifications \cite{Weinberg:1966jm,Klauder:1969zz,Wilson:1994fk,Brodsky:1997de,Hiller:2016itl} compared to standard equal-time (ET) quantization, where one uses surfaces of constant Lorentzian time.  One pays a conceptual price for this simplification, however.  Important physics effects, such as spontaneous symmetry breaking and  renormalization of the vacuum energy, are subtle to uncover in LC quantization \cite{Chang:1968bh,Yan:1973qg,Maskawa:1975ky,Tsujimaru:1997jt,Yamawaki:1998cy,Heinzl:2003jy,Beane:2013ksa,Herrmann:2015dqa,Collins:2018aqt,Martinovic:2018apr}.  Many of the difficult subtleties of LC can be traced to the fact that energy $p_+ = \frac{\mu^2}{2p_-}$ is inversely proportional to momentum $p_-$ in terms of the Lorentz-invariant $\mu^2 = p^2$, and consequently ``zero modes'' with vanishing $p_-$ have infinite LC energy and are lifted out of the spectrum.

From an Effective Field Theory (EFT) perspective, the heavy zero modes must be integrated out, potentially leaving behind new interactions compared to the theory in ET quantization.  It is useful to think in terms of an ``effective lightcone Hamiltonian'' $H_{\rm eff}$ for the lightcone-quantized theory, containing any new interactions generated by integrating out the LC zero modes.  A general prescription for how to obtain $H_{\rm eff}$ starting from an ET Hamiltonian $H_{\rm ET}$ is an open problem.  In \cite{US}, we proposed a method for constructing $H_{\rm eff}$ in terms of $H_{\rm ET}$ perturbatively, but ultimately one would like to go beyond perturbation theory.  At a minimum, one needs to be able to determine which relevant and marginal operators appear in $H_{\rm eff}$, at which point their coefficients can be fixed in principle in terms of physical observables.  More ambitiously, one would like to be able to nonperturbatively determine {\it a priori} the values of the bare parameters in $H_{\rm eff}$.

A useful model for investigating these issues in detail is $\lambda \phi^4$ scalar theory in $d=2$.  In this case, the perturbative prescription in \cite{US} reduces to the earlier prescription of \cite{Burkardt,Burkardt2}, which says that the entire effect of the zero modes is simply a shift in the bare parameters:
\be
m^2 \rightarrow m^2_{\rm eff} = m^2 + 12\lambda \< \phi^2\>.
\label{eq:Burkardt}
\ee
As we will review, this prescription passes nontrivial checks at the perturbative level, but fails nonperturbatively.  However, comparisons of numerical analyses of the theory in LC quantization indicate that there is a critical value of the mass, or more precisely of the dimensionless ratio $\bar{\lambda} = \frac{\lambda}{m^2}$, where the theory reaches a scale-invariant fixed point in the IR, but with a shifted value of the critical coupling as compared to numerical analyses of ET quantization.  So, one may take this as evidence that although the exact form (\ref{eq:Burkardt}) for $m_{\rm eff}^2$ is only valid perturbatively, there is still some value of $m_{\rm eff}^2$ that matches the ET theory.  In other words, the Lagrangian in ET and LC quantization are really describing the same theory, once their respective bare parameters have been fixed in terms of a physical observable.  In this paper, we will argue that this is true, and that furthermore one can extract the correct matching of bare parameters from the perturbative equation (\ref{eq:Burkardt}), though in a more involved way than one might naively have expected.  

The basic idea is that the physical mass gap $\mu_{\rm gap}$ in the theory should be the Borel resummation of its perturbation series in both ET and LC quantization.  This is the main assumption of the procedure we apply for extracting a map between bare parameters of the two quantizations.  Assuming this is true, then (\ref{eq:Burkardt}) relates the two perturbation series to all orders, and therefore together with Borel resummation it allows one to calculate $\mu_{\rm gap}$ as a function of the bare parameters in both quantization schemes starting with just the perturbation series of $\mu_{\rm gap}$ and $\<\phi^2\>$ in ET.  Since the gap is a physical quantity, one can then extract a map between bare parameters by equating the gap obtained as a function of $\bar{\lambda}$ in LC  and in  ET. 

To implement this procedure, we will use the recent results of \cite{Serone} obtaining the perturbation series of the gap and the vacuum energy to eighth order in the coupling.  We will also closely follow their implementation of Borel resummation, originally from \cite{ZinnJustin}, which involves optimizing over two additional parameters to improve the convergence rate.  We reproduce their Borel resummation of the mass gap in ET quantization, and additionally obtain the Borel resummation of the mass gap in LC quantization.  From this calculation of the gap, we can extract a map $\bar{\lambda}_{LC}(\bar{\lambda}_{\rm ET})$.  This result is shown in Fig.~\ref{fig:Map}. 

In principle, with high enough orders in the perturbation series, one should get the same result independently of whether one Borel resums the perturbation series for the mass gap or some power $\mu_{\rm gap}^\alpha$.  In practice, with only finitely many perturbative terms, the result does depend on which power of the gap one chooses to resum.  In ET quantization, the convergence rate is fastest if one Borel resums $\mu_{\rm gap}$ \cite{Serone}, due to the fact that the critical exponent $\nu=1$ in $d=2$ and therefore $\mu_{\rm gap} \sim |\lb-\lb_*|$ as one approaches the critical coupling $\lb_*$. By contrast, in LC, we argue that $\mu_{\rm gap}^2$ should close linearly in the bare coupling as one approaches the critical point. Therefore, we expect the convergence rate to be fastest if we Borel resum $\mu_{\rm gap}^2$ in LC quantization.

To test our procedure, we compare the results of the Borel resummation to physical quantities calculated in ET and LC quantization using the nonperturbative methods of Hamiltonian truncation (i.e.~ET renormalized Hamiltonian truncation \cite{Yurov:1989yu,Yurov:1991my,Coser:2014lla,Hogervorst:2014rta,Rychkov:2014eea,Rychkov:2015vap,Elias-Miro:2015bqk,Bajnok:2015bgw,Elias-Miro:2017xxf,Elias-Miro:2017tup,Hogervorst:2018otc} and LC conformal truncation \cite{Katz:2013qua,Katz:2014uoa,Katz:2016hxp,Anand:2017yij,US,Delacretaz:2018xbn}, respectively).  That is, we use Hamiltonian truncation to numerically compute physical quantities as a function of the bare parameters in ET and LC, and then use the map $\bar{\lambda}_{\rm LC}(\bar{\lambda}_{\rm ET})$ obtained from Borel resummation to compare them.  The first physical quantity we compare is just the mass gap itself, and we find very good agreement, as shown in Fig.~\ref{fig:Comparison}.  Because the mass gap is the quantity that we used to extract the map between parameters, this test is equivalent to a check that Borel resumming the gap works well in both ET and LC.  

The second physical quantity that we compare is the residue $Z$ of the single-particle pole of the $\phi$ propagator.  Equivalently, it is (the square of) the matrix element of $\phi$ between the vacuum and the single-particle state.  We find that the ET and LC results for $Z$ in terms of the physical quantity $\mu_{\rm gap}^2/\lambda$ agree over a wide range of couplings until close to the critical point, where truncation effects limit the convergence rate. This agreement for $Z$  is further evidence that both ET and LC quantization compute the same physical observables once we identify the bare parameters with our matching procedure.

The rest of the paper is organized as follows.  In section \ref{sec:PertMatch}, we review the perturbative matching between ET and LC, and how its naive extension to a nonperturbative matching fails. In section \ref{sec:map}, we obtain the matching between bare parameters by Borel resumming the mass gap both as a function of ET parameters and LC parameters, and equating them.  In section  \ref{sec:Tests}, we perform tests of the mapping by using it to compare physical quantities computed with conformal truncation techniques in the two quantizations. Finally, in section \ref{sec:future}, we conclude with a discussion of potential future directions.

\section{Review of Perturbative Matching}
\label{sec:PertMatch}

In this section, we will review the perturbative matching between the bare parameters in $\lambda \phi^4$ theory in LC quantization vs ET quantization.  We will also discuss the difficulty in extending the perturbative matching to the nonperturbative level.

\subsection{Shift in Bare Parameters}

We consider the following Lagrangian in $d=2$:
\be
\CL = \frac{1}{2} \norder{(\partial \phi)^2} - \frac{1}{2} m^2 \norder{\phi^2} -  \lambda \norder{\phi^4},
\label{eq:Phi4L}
\ee
where $\norder{\Ocal}$ indicates that the operator is normal-ordered. The theory has a single dimensionless parameter that we will denote $\bar{\lambda} \equiv \frac{\lambda}{m^2}$.\footnote{Note that, due to normal-ordering, the bare mass parameter $m^2$ is finite.}  

\begin{figure}[t!]
\begin{center}
\includegraphics[width=0.4\textwidth]{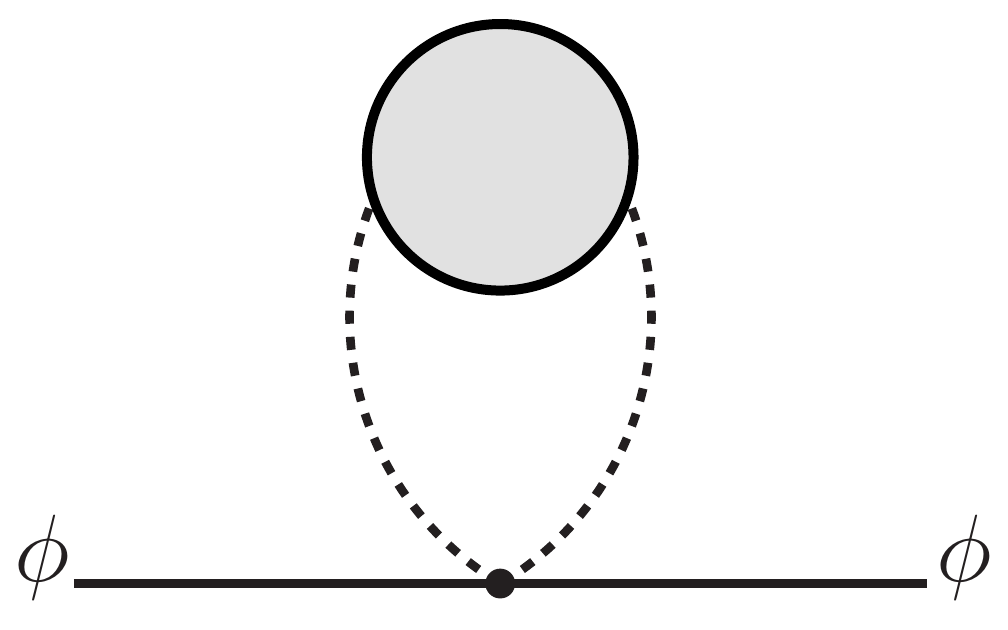}
\caption{General structure of ``plant'' diagrams.}
\label{fig:Plant}
\end{center}
\end{figure}

The proposal in \cite{US} for how to determine the effective LC Hamiltonian $H_{\rm eff}$ is essentially a matching procedure, where correlators are computed in ET and LC quantization, and new terms are added to the LC Hamiltonian to make them agree.  More explicitly, this matching is achieved using the following trick.  The matrix elements of the ET Hamiltonian can be read off from matrix elements of the unitary evolution operator $U(t)$ through the relation $H = \lim_{t\rightarrow 0} i \partial_t U(t)$.  Matrix elements of $U(t)$ are simply given by two-point functions of operators, which are independent of the quantization scheme.  However, the LC Hamiltonian generates evolution with respect to $x^+$ rather than $t$, so we extract it from $U(t)$ by taking
\be
H_{\rm eff} = \lim_{x^+ \rightarrow 0} i \partial_{x^+} U(x^+),
\ee
where the partial derivative is now taken with respect to $x^+$ rather than $t$.  The spatial coordinates are also treated differently: the external states in ET have fixed momentum $P_x$, whereas in LC they have fixed lightcone momentum $P_-$, so in the former case we Fourier transform with respect to $x$ and in the latter with respect to $x^-$. Perturbatively, one can evaluate $U$ in terms of its Dyson series.  Naively, only the linear term in $\lambda$ in the Dyson series contributes, both for $H$ and for $H_{\rm eff}$, because higher order terms involve multiple integrals over time, all of whose region of integration vanishes in the limit $t\rightarrow 0$ or $x^+ \rightarrow 0$, respectively.  The subtlety is that in LC coordinates, the higher order terms in the Dyson series can contain $\delta$ functions of LC time, which therefore can have a nonvanishing contribution even from an infinitesimal region of integration.\footnote{See also \cite{Chabysheva:2018wxr} for another perspective on why interpolations between LC and ET are discontinuous at the LC limit.} Such $\delta$ functions in position space correspond to contributions independent of some momentum $q_+$ flowing through the diagram in momentum space, or more generally, to contributions that are simply polynomials in $q_+$.  In $\lambda \phi^4$ theory, the class of diagrams that depend on $q_+$ this way are diagrams with the topology of a ``plant'' shown in Fig.~\ref{fig:Plant}, i.e. the diagram is an arbitrarily complicated subdiagram connected to a scalar line at a single point.\footnote{The basic idea, explained more thoroughly in \cite{US}, is that most diagrams have $q_+$ dependence in the denominator $\sim \frac{i}{2 q_+ q_- -m^2+i \epsilon}$ of internal propagators.  For plant diagrams, however, none of the external spatial momentum $p_-$ flows through the nontrivial part of the diagram, so there is a contribution from the region of integration where the loop momentum $q_-$ vanishes and therefore the denominator does not depend on $q_+$.}  This conclusion reproduces an earlier result due to Burkardt  \cite{Burkardt}, from inspection of Feynman diagrams.  It is clear that in perturbation theory, these plant diagrams simply renormalize the mass by a shift proportional to the loop diagrams for the vev (in ET quantization) of $\norder{\phi^2}$,
\be
m^2_{\rm LC} = m_{\rm ET}^2 + 12\lambda_{\rm ET}\< \norder{\phi^2}\>.
\label{eq:PertMatching}
\ee

Remarkably, in a heroic effort, the perturbative coefficients of both the vacuum energy density $\Lambda(\lb)$ and the mass gap $\mu_\gap(\lb)$ have recently been computed to $\CO(\lb^8)$ in this theory \cite{Serone}.  We summarize their result here:
\be
\bar{\Lambda}_{\rm ET} \equiv \frac{\Lambda}{m_{\rm ET}^2} = \sum_{n=2}^\infty a_n \bar{\lambda}^n_{\rm ET}, \quad
\bar{\mu}^2_{\rm ET} \equiv \frac{\mu_{\rm gap}^2}{m_{\rm ET}^2} = 1+ \sum_{n=2}^\infty c_n \bar{\lambda}^n_{\rm ET},  
\ee
where
\be
&& a_2= -\frac{21 \zeta(3)}{16 \pi^3}, \, a_3= \frac{27 \zeta(3)}{8 \pi^4}, \, a_4= - 0.11612596491, \, a_5= 0.394953418, \nn\\
 && a_6=-1.62979422, \, a_7= 7.8540421, \, a_8=-43.192021, \\
 && c_2 =  - \frac{3}{2} , \, c_3 = \frac{9}{\pi} + \frac{63 \zeta(3)}{2 \pi^3}, \, c_4 = -14.65586922, \, c_5=65.9730843, \nn\\
 && c_6 = -347.888128, \, c_7 = 2077.70336, \, c_8 = -13\ 711.0454 .
\ee
We can extract the vev $\< \norder{\phi^2}\>$ from the vacuum energy by taking derivatives.  Naively, since the coefficient of $\norder{\phi^2}$ in the action is $m^2$, the derivative of the vacuum energy with respect to $m^2$ is $\<\norder{\phi^2}\>$.  However, $m^2$ also shows up in the action through the renormalization scheme. Specifically, in defining the Lagrangian~\eqref{eq:Phi4L} in terms of normal-ordered operators, we subtracted off the divergent contributions to the vacuum energy and bare mass, which depend on $m^2$. We can make this additional $m^2$-dependence explicit by rewriting the Lagrangian without normal-ordering, using the relations~\cite{Rychkov:2014eea,Rychkov:2015vap}
\bq
\norder{\phi^2} = \phi^2 - \fr{1}{4\pi}\log\fr{\Lambda_{\rm cutoff}^2}{m^2}, \quad
\norder{\phi^4} = \phi^4 - \fr{3}{2\pi}\log\fr{\Lambda_{\rm cutoff}^2}{m^2} \phi^2 + \fr{3}{16\pi^2} \log^2\fr{\Lambda_{\rm cutoff}^2}{m^2},
\eq
where we've imposed a uniform cutoff $\Lambda_{\rm cutoff}$ on the loop momenta. The resulting expression for the Lagrangian is
\be
\CL &=& \frac{1}{2} (\partial \phi)^2 - \frac{1}{2} \Big(m^2-\fr{3\lambda}{\pi} \log \fr{\Lambda_{\rm cutoff}^2}{m^2}\Big) \phi^2 - \lambda \phi^4 -\delta \Lambda, \\
\frac{\partial \delta \Lambda}{\partial m^2}  &=& -\fr{1}{8\pi}   \left(1 + \frac{3 \bar{\lambda}}{\pi} \right) \log \fr{\Lambda_{\rm cutoff}^2}{m^2} ,
\ee
where we have added mass and vacuum energy counterterms. 
So the actual relation between the vacuum energy and the vev $\< \norder{\phi^2}\>$ is
\be
\frac{\partial}{\partial m^2} \Lambda = \half \bigg(1 + \fr{3\bar{\lambda}}{\pi}\bigg) \bigg( \<\phi^2\> - \fr{1}{4\pi} \log \fr{\Lambda_{\rm cutoff}^2}{m^2} \bigg) = \half \bigg(1 + \fr{3\bar{\lambda}}{\pi}\bigg) \<\norder{\phi^2}\>.
\ee
In terms of the dimensionless quantities $\bar{\Lambda}, \bar{\lambda}$, this equation takes the following form, which is what we will use to extract $\< \norder{\phi^2}\>$:
\be
\< \norder{\phi^2}\> = 2 \frac{\bar{\Lambda}- \bar{\lambda}\frac{d}{d\bar{\lambda}} \bar{\Lambda}}{1+ \frac{3 \bar{\lambda}}{\pi}} .
\label{eq:VevDef}
\ee 
From now on we will suppress the normal-ordering notation $\norder{\Ocal}$, with the understanding that all local operators are to be normal-ordered.

Once we have the perturbative expansion of the vev $\< \phi^2\>$, we can obtain the perturbative relation between the LC and ET couplings $\bar{\lambda}$.  Equation (\ref{eq:PertMatching}) implies that
\be
\bar{\lambda}_{\rm LC} &=& \frac{\lambda}{m_{\rm LC}^2} = \frac{\bar{\lambda}_{\rm ET}}{1 + 12 \bar{\lambda}_{\rm ET} \< \phi^2\>}.
\label{eq:lambdaET2LC}
\ee 
Using the perturbative coefficients of $\Lambda$, the first few coefficients of $\bar{\lambda}_{\rm LC}$ in terms of $\bar{\lambda}_{\rm ET}$ are
\be
\bar{\lambda}_{\rm LC} = \bar{\lambda}_{\rm ET} - \frac{63 \zeta(3)}{2\pi^3} \bar{\lambda}_{\rm ET}^4 + \frac{513 \zeta(3) }{2\pi^4} \bar{\lambda}_{\rm ET}^5 + \dots .
\ee
It is straightforward to invert this equation to any order in perturbation theory:
\be
\bar{\lambda}_{\rm ET} = \bar{\lambda}_{\rm LC} + \frac{63 \zeta(3)}{2\pi^3} \bar{\lambda}_{\rm LC}^4 - \frac{513 \zeta(3)}{2 \pi^4} \bar{\lambda}_{\rm LC}^5 + \dots. 
\ee
To compare the gap $\bar{\mu}_{\rm ET}$ to a LC calculation, we divide the gap by the LC parameter $m_{\rm LC}$ and express the result in terms of $\bar{\lambda}_{\rm LC}$:
\be
\bar{\mu}^2_{\rm LC}(\bar{\lambda}_{\rm LC}) = \frac{\mu_{\rm gap}^2}{m_{\rm LC}^2}  = \frac{\bar{\mu}_{\rm ET}^2(\bar{\lambda}_{\rm ET})}{1+12 \bar{\lambda}_{\rm ET} \< \phi^2\>} ,
\label{eq:muET2LC}
\ee
where $\bar{\lambda}_{\rm ET}$ is converted to a function of $\bar{\lambda}_{\rm LC}$ by inverting (\ref{eq:lambdaET2LC}).\footnote{Equivalently, $
\bar{\mu}^2_{\rm ET}(\lambda_{\rm ET}) = (1+12 \bar{\lambda}_{\rm ET} \< \phi^2\>) \bar{\mu}^2_{\rm LC} \left( \frac{\bar{\lambda}_{\rm ET}}{1+12 \bar{\lambda}_{\rm ET} \< \phi^2 \> } \right).$
}
Expanded out to $\bar{\lambda}_{\rm LC}^6$, the prediction for the gap in  LC  quantization is
\be
\bar{\mu}^2_{\rm LC} &=& 1 - \frac{3}{2} \bar{\lambda}^2_{\rm LC} + \frac{9}{\pi} \bar{\lambda}^3_{\rm LC} - 11.4906 \bar{\lambda}^4_{\rm LC} + 52.7576 \bar{\lambda}_{\rm LC}^5 - 287.357 \bar{\lambda}_{\rm LC}^6 + \dots .
\label{eq:NumLCCoeffs}
\ee
We have independently computed these coefficients up to $\bar{\lambda}_{\rm LC}^5$ in LC quantization using old-fashioned perturbation theory.  More precisely, we computed the Hamiltonian in LC quantization in a basis of operators with dimension up to $\Delta_{\rm max}$, and then we substituted these matrix elements into the time-independent perturbation theory formula for the single-particle state energy.\footnote{We also had to extrapolate our results to infinite $\Delta_{\rm max}$, since we were limited by computation time to $\Delta_{\rm max}\le 33$. We extrapolated by fitting the dependence of each perturbative coefficients on $\Delta_{\rm max}$  with a power law, $ a \Delta_{\rm max}^{-N} + b$,  where $a,b$ and $N$ were obtained from fitting.  The main source of error on the coefficients is due to uncertainties in the fit parameters $a,b,$ and $N$; we estimate that this error is in the last digit shown in each coefficient in (\ref{eq:NumLCCoeffs}).}  We obtained the numeric result
\be
\bar{\mu}^2_{\rm LC} &=& 1- 1.49995 \bar{\lambda}^2_{\rm LC} + \frac{8.9999}{\pi} \bar{\lambda}^3_{\rm LC} - 11.52 \bar{\lambda}^4_{\rm LC} + 52.9 \bar{\lambda}_{\rm LC}^5+ \dots,
\ee
in reasonable agreement with (\ref{eq:NumLCCoeffs}).

\subsection{Nonperturbative Failure}

Next, we would like to generalize the perturbative matching condition (\ref{eq:PertMatching}) to a nonperturbative relation.  The most natural guess would be that (\ref{eq:PertMatching}) is simply true exactly, giving $m_{\rm LC}$ directly as a function of $\lambda_{\rm ET}$ once the nonperturbative vev $\< \phi^2\>$ is known as a function of $\lambda_{\rm ET}$.  However, as noted in \cite{US}, this guess is not consistent with numeric results obtained using Hamiltonian truncation, or with results from Borel resummation~\cite{Serone}.  We will review the relevant numeric results here.

\begin{figure}[t!]
\begin{center}
\includegraphics[width=0.42\textwidth]{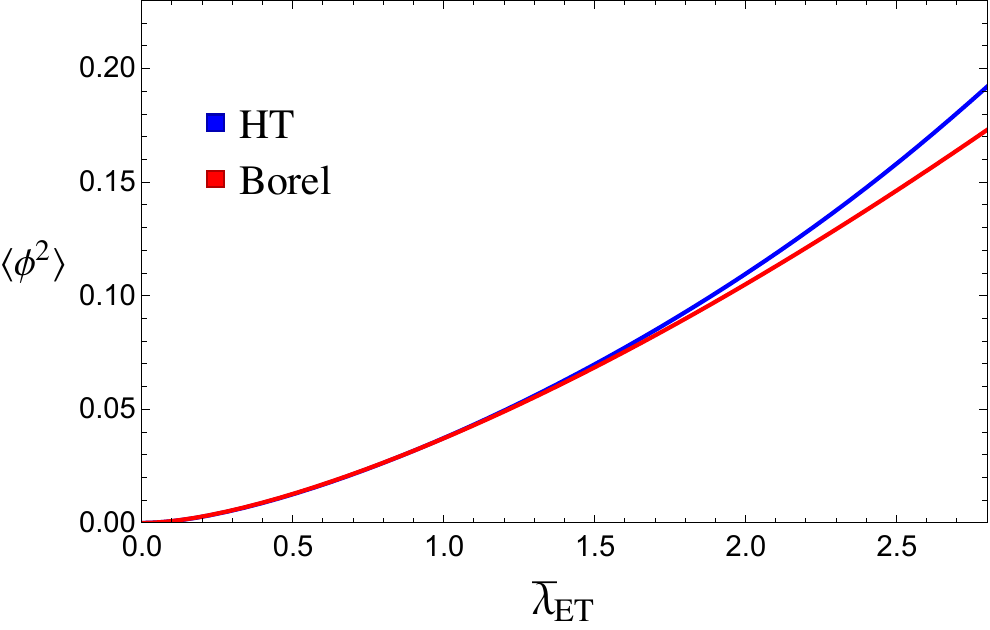}
\includegraphics[width=0.42\textwidth]{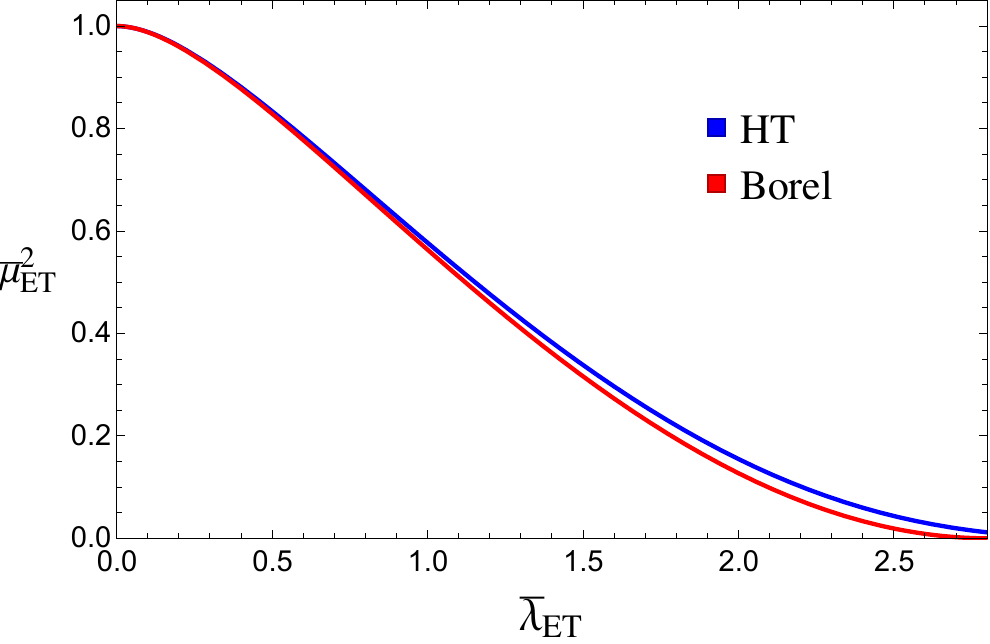}
\includegraphics[width=0.42\textwidth]{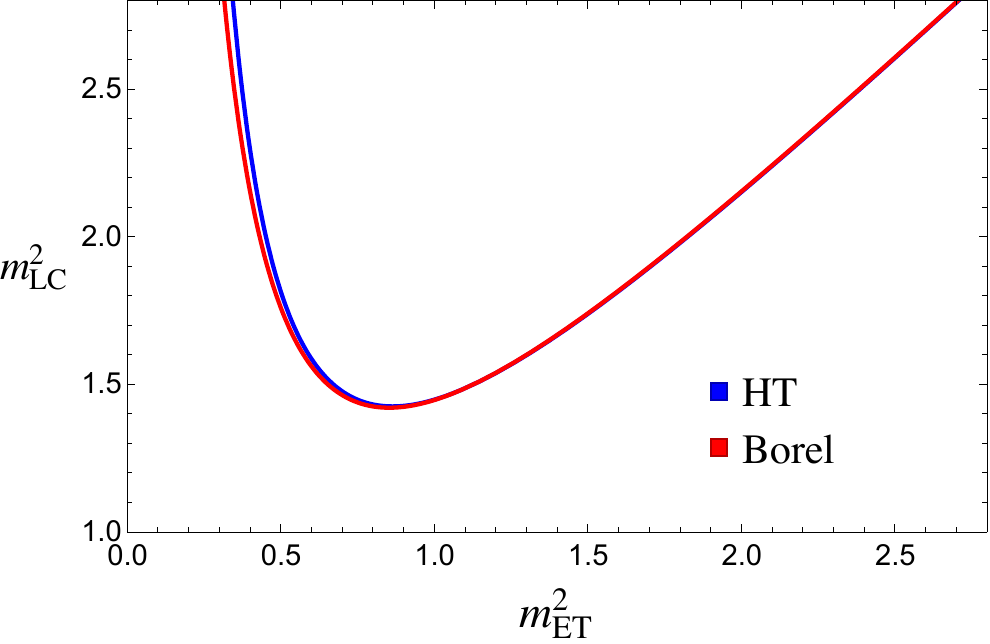}
\includegraphics[width=0.42\textwidth]{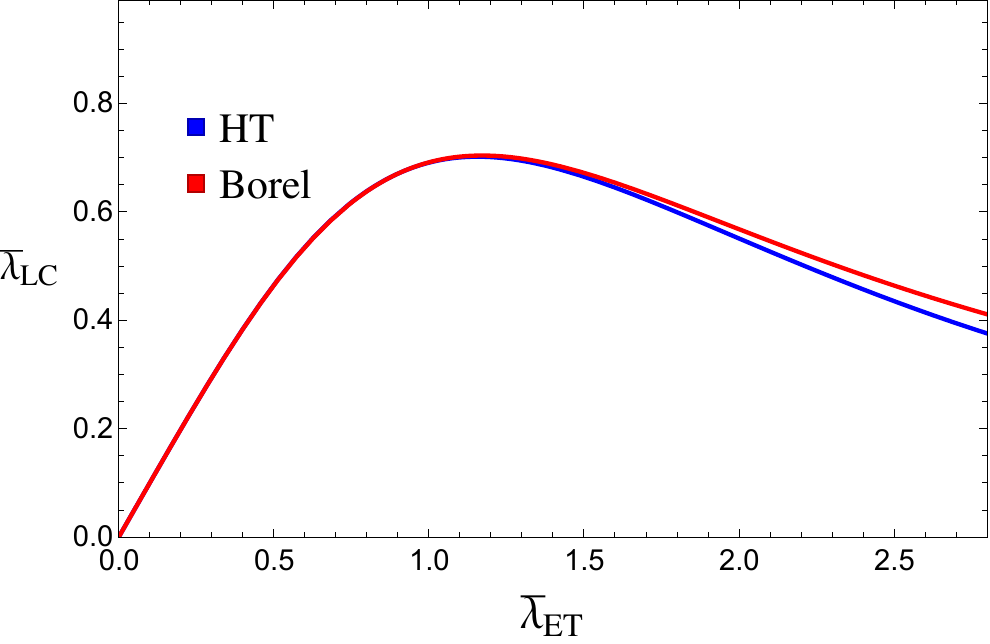}
\includegraphics[width=0.47\textwidth]{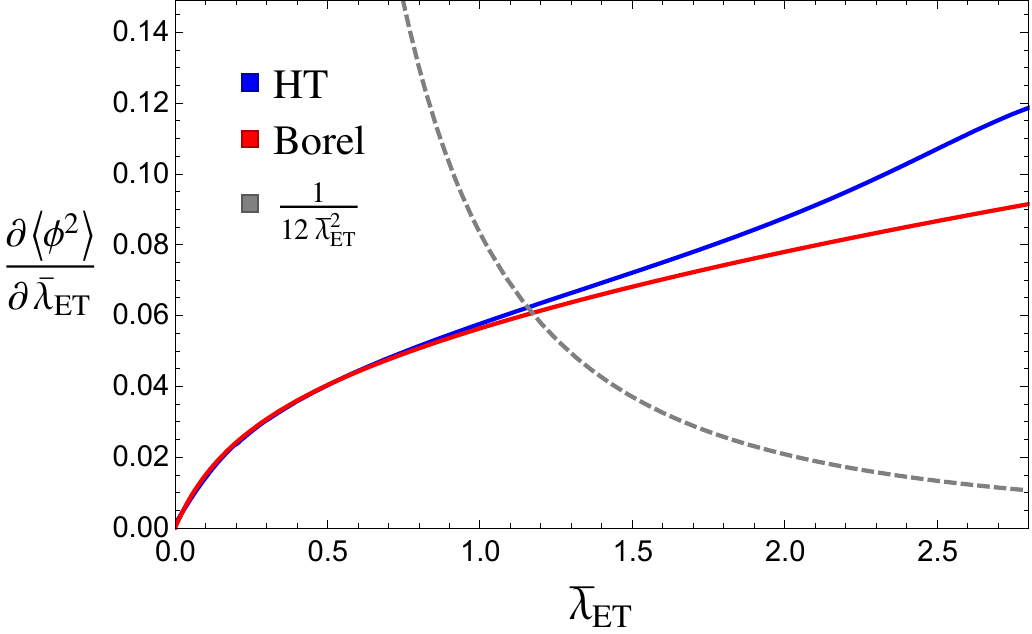}
\caption{{\it Left, top}: Plot of $\< \phi^2\>$ as a function of $\bar{\lambda}_{\rm ET}$. {\it Right, top}: Mass-squared gap $\bar{\mu}^2_{\rm ET}$ in ET quantization as a function of $\bar{\lambda}_{\rm ET}$. {\it Left, middle:} $m^2_{\rm LC}$ as a function of $m^2_{\rm ET}$, setting $\lambda_{\rm ET}= \lambda_{\rm LC} =1$, according to eq.~(\ref{eq:PertMatching}). {\it Right, middle:}   $\bar{\lambda}_{\rm LC}$ as a function of $\bar{\lambda}_{\rm ET}$, again according to eq.~(\ref{eq:PertMatching}). {\it Bottom:}   $\fr{\p\<\phi^2\>}{\p \bar{\lambda}_{\rm ET}}$ (solid black line) compared with the turnaround threshold $\fr{1}{12\bar{\lambda}_{\ET}^2}$ (dashed gray line). All five plots include results obtained using both renormalized ET Hamiltonian truncation~\cite{Rychkov:2014eea} (``HT'') and Borel resummation~\cite{Serone} (``Borel''). 
The close agreement between the HT and Borel methods is evidence of their accuracy. The turnaround in the middle two plots indicate that the literal interpretation of (\ref{eq:PertMatching}) would incorrectly imply that the map from $\bar{\lambda}_{\rm ET}$ to $\bar{\lambda}_{\rm LC}$ is not invertible; two different values of $\bar{\lambda}_{\rm ET}$ would correspond to the same $\bar{\lambda}_{\rm LC}$. }
\label{fig:NonPertBurkFail}
\end{center}
\end{figure}

To test the conjecture that (\ref{eq:PertMatching}) is true as an exact statement, we can take $\bar{\mu}^2_{\rm ET}$ and $\< \phi^2\>$ from a numeric computation in ET quantization as a function of $\bar{\lambda}_{\rm ET}$, as well as $\bar{\mu}^2_{\rm LC}$ numerically in LC quantization as a function of $\bar{\lambda}_{\rm LC}$, and use eqs.~(\ref{eq:lambdaET2LC}) and (\ref{eq:muET2LC}) to convert $\bar{\mu}^2_{\rm ET}(\bar{\lambda}_{\rm ET})$ to $\bar{\mu}^2_{\rm LC}(\bar{\lambda}_{\rm LC})$.  The result of the numeric computation of $\< \phi^2\>$ is shown in Fig.~\ref{fig:NonPertBurkFail}.\footnote{Concretely, the nonperturbative ET data for $\<\phi^2\>$ was obtained via eq.~\eqref{eq:VevDef} from the vacuum energy $\Lambda$ initially computed with renormalized Hamiltonian truncation in~\cite{Rychkov:2014eea}, as well as the computation of $\Lambda$ via Borel resummation in~\cite{Serone}.} Immediately, however, one encounters a problem.  The issue is that with the vev $\<\phi^2\>$ as shown, the map (\ref{eq:lambdaET2LC}) from $\bar{\lambda}_{\rm ET}$ hits a local maximum at around $\bar{\lambda}_{\rm LC} \approx 0.7$ and then turns around.  If this prediction were correct, it would mean that no value of $\bar{\lambda}_{\rm ET}$ would correspond to $\bar{\lambda}_{\rm LC} \gtrsim 0.7$. Equally problematically, it would imply that a single value of $\bar{\lambda}_{\rm ET}$ would correspond to two different values of $\bar{\lambda}_{\rm LC}$. Neither of these bizarre predictions is seen in the numeric analysis of LC quantization, as we review below.

We emphasize that this turnaround problem does not depend on any particularly special feature of the numeric result for $\< \phi^2\>$.  From (\ref{eq:lambdaET2LC}), it is easy to see that such a turnaround occurs if at any point
\be
\frac{d}{d\bar{\lambda}_{\rm ET}} \< \phi^2\> = \frac{1}{12 \bar{\lambda}_{\rm ET}^2}.
\ee
Since $\frac{d}{d \bar{\lambda}_{\rm ET}} \< \phi^2\>$ starts out small at small $\bar{\lambda}_{\rm ET}$, its derivative must therefore stay below $(12 \bar{\lambda}_{\rm ET}^2)^{-1}$ to avoid a turnaround.  Therefore, even rather modest growth in $\< \phi^2\>$ as a function of coupling leads to the above problem eventually.

Less obvious a priori is the fact that the turnaround point occurs at smaller values of the coupling than the critical coupling $\bar{\lambda}_{\rm ET,*}$, where the gap closes.  A reasonable conjecture would have been that (\ref{eq:PertMatching}) is valid nonperturbatively but only in the unbroken phase.  However, one can see from Fig.~\ref{fig:NonPertBurkFail} that the turnaround occurs for $\bar{\lambda}_{\rm ET} \sim 1$, whereas the critical point is at the much larger value $\bar{\lambda}_{\rm ET} \sim 3$, so this conjecture is also wrong. It appears that if (\ref{eq:PertMatching}) has some nonperturbative meaning, it must be more subtle.  A hint is that the physical quantity $\bar{\mu}^2_{\rm ET}$ is an asymptotic series of the coupling, but one that Borel resums to the true value \cite{Serone}.  So in principle, $\bar{\mu}^2_{\rm ET}$ is determined by its perturbation series through Borel resummation, and one might hope this is true of $\bar{\mu}^2_{\rm LC}$ in terms of $\bar{\lambda}_{\rm LC}$ as well.  Then, the perturbative equation (\ref{eq:PertMatching}) would simply be the connection between the two perturbation series, and would need to be combined with Borel summation to obtain a nonperturbative matching relation between the two quantizations.  In the following sections, we will turn to analyzing this possibility in detail.  We relegate to appendix \ref{sec:0dexample} a discussion of zero-dimensional analogue, where one can see more explicitly how a relation like (\ref{eq:PertMatching}) might have a straightforward interpretation to all orders in perturbation theory, but involve additional subtleties nonperturbatively.

\section{Map from ET to LC Using Borel Resummation of Mass Gap}
\label{sec:map}

We've now seen that the proposed map \eqref{eq:PertMatching} between ET and LC couplings, which holds to all orders in perturbation theory, clearly fails nonperturbatively. Based on this result, one might na\"{i}vely suspect that no such nonperturbative map exists, or at least cannot be found by knowing only perturbative data. However, in this section, we demonstrate that the map $\lb_\LC(\lb_\ET)$ \emph{can} be constructed by instead computing the mass gap $\mb_\gap^2(\lb)$ in both ET and LC quantization by Borel resumming the two perturbation series, then matching the two functions to indirectly obtain the nonperturbative map between the two couplings,
\be
\mu_{\rm gap, LC} = \mu_{\rm gap, ET} \leftrightarrow \fr{\mb_\gap^2(\lb_\LC)}{\lb_\LC} = \fr{\mb_\gap^2(\lb_\ET)}{\lb_\ET} \quad \Rightarrow \quad \lb_\LC(\lb_\ET).
\ee

The fact that the intermediate function $\mb_\gap^2(\lb)$ can be accurately computed by Borel resumming its perturbative expansion was demonstrated recently for the case of ET quantization in \cite{Serone}. There, the authors directly computed the perturbative expansion of $\mb_\gap^2(\lb_\ET)$ up to $O(\lb^8_\ET)$, then used these series coefficients to numerically determine the fully resummed function. These resummation results successfully reproduced previous nonperturbative calculations of $\mb_\gap^2(\lb_\ET)$ via Hamiltonian truncation.

In this section, we repeat this procedure for the case of LC quantization. Specifically, we use eq.~\eqref{eq:PertMatching} to convert the perturbative expansion of $\mb_\gap^2$ in powers of $\lb_\ET$ into the corresponding expansion in terms of $\lb_\LC$. Using the same approach as~\cite{Serone}, we then use these new LC perturbative coefficients to numerically determine the resummed function $\mb_\gap^2(\lb_\LC)$. Once we have this function, we can combine it with the results of \cite{Serone} to finally obtain the desired nonperturbative map $\lb_\LC(\lb_\ET)$.\footnote{Note that we are not simply Borel resumming the perturbative expansion of $\bar{\lambda}_{\rm LC}(\bar{\lambda}_{\rm ET})$ from (\ref{eq:PertMatching}), since for $\bar{\lambda}_\ET$ less than the turnaround point, that should just reproduce the naive prescription where we apply (\ref{eq:PertMatching}) as an exact relation. For larger $\bar{\lambda}_\ET$, the Borel integral should diverge, since it is attempting to reproduce a noninvertible function.}

\begin{figure}[t!]
\begin{center}
\includegraphics[width=0.65\textwidth]{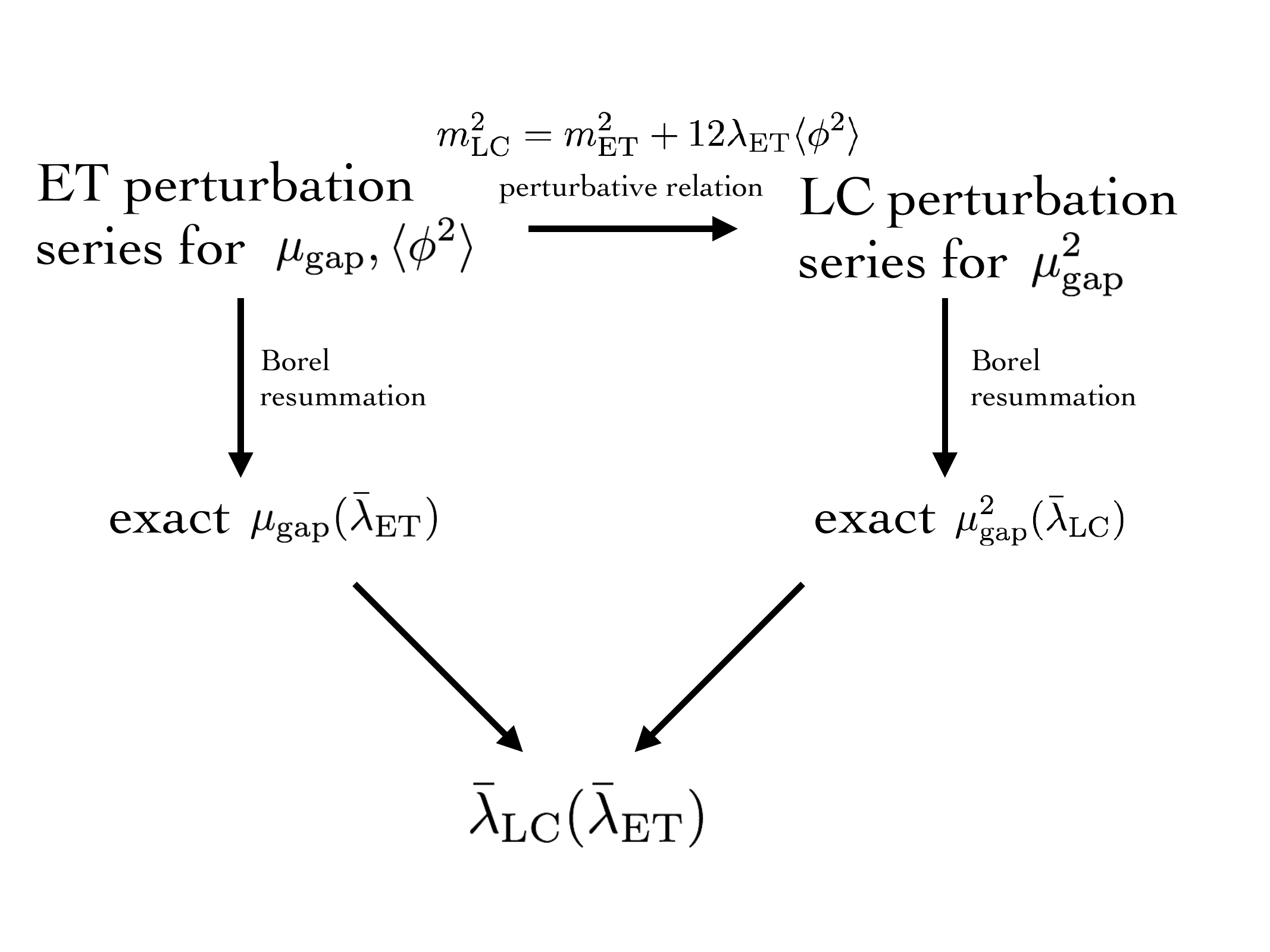}
\caption{Outline of the procedure for extracting the map between bare couplings in LC and ET from the ET perturbation series.}
\label{fig:sketch}
\end{center}
\end{figure}

It is worth emphasizing that in this entire calculation, \emph{we only use data obtained in ET quantization}. The perturbative expansion of $\mb_\gap^2(\lb_\LC)$ is obtained solely from the ET expansions for $\mb_\gap^2$ and $\<\phi^2\>$, combined with the perturbative map \eqref{eq:PertMatching}. This strategy is sketched in Fig.~\ref{fig:sketch}.  In section~\ref{sec:Tests}, we compare our resummation results with Hamiltonian truncation results obtained directly in LC quantization, but at this stage we are using strictly ET data.

\subsection{Lightning Review of Borel Resummation via Conformal Mapping}

Before focusing on the application to LC quantization, let's briefly review the resummation technique used in \cite{Serone}, though interested readers should consult that work for more details. In general, we are interested in studying a function $F(g)$, which has the asymptotic expansion
\be
F(g) = \sum_{n=0}^\infty F_n \, g^n,
\ee
where, for our particular case of interest, $F\ra \bar{\mu}_\gap^2$ and $g \ra \lb$. In principle, we would like to compute the Borel-Le Roy transform
\be
\Bcal_b(t) = \sum_{n=0}^\infty B^{(b)}_n t^n, \qquad B^{(b)}_n \equiv \fr{F_n}{\G(n+b+1)},
\ee
which can then be used to obtain the Borel resummed function
\be
F_B(g) = \fr{1}{g^{b+1}} \int_0^\infty dt \, t^b e^{-t/g} \Bcal_b(t).
\ee
However, we do not have the full asymptotic series, only the first $N+1$ terms,
\be
F^{(N)}(g) = \sum_{n=0}^N F_n \, g^n,
\ee
and if we na\"{i}vely apply this procedure to the truncated sum, we simply get the same expression back. The problem is that the Borel transform $\Bcal_b(t)$ only has a finite radius of convergence, due to a singularity at $t=-\fr{1}{a}$,\footnote{The location of this singularity is given by the classical action of the leading instanton configuration, which can be computed numerically to obtain $a = 0.683708$~\cite{Serone}. Under mild assumptions about the series coefficients of the gap and vacuum energy in ET, the same $a$ should control the asymptotic behavior of the LC series coefficients.  We have also explicitly checked that the difference in ratios of subsequent terms for ET and LC coefficients is only a couple percent at  eighth order.} but the inverse Borel transform evaluates $\Bcal_b(t)$ far beyond this radius of convergence. Because of this, we cannot exchange the sum with the integration and inverse Borel transform term-by-term.

We can avoid this issue with the change of variables,
\be
t = \fr{4}{a} \fr{u}{(1-u)^2},
\ee
which maps the entire complex plane to the unit disk $|u| \leq 1$, with the branch cut singularity originating from $t = -\fr{1}{a}$ mapped to the edge of the disk. If we now study the series expansion of $\tB_b(u) \equiv \Bcal_b(t(u))$, we find that it converges over the entire range of integration, which means we can safely inverse Borel transform each term in the sum,
\be
F_B(g) = \fr{1}{g^{b+1}} \sum_{n=0}^\infty \widetilde{B}_n^{(b)} \int_0^1 du \fr{dt}{du} t^b(u) \, e^{-t(u)/g} u^n.
\ee
Since the coefficients $\widetilde{B}_n^{(b)}$ only depend on $F_{n'}$ with $n' \leq n$, truncating the original asymptotic series to the first $N+1$ terms is equivalent to restricting this sum for $F_B(g)$ to its first $N+1$ terms. However, this new series is a \emph{convergent} one, which makes this truncated sum a reasonable approximation to the full expression.

We can actually improve the convergence of this series by introducing a second parameter $s$, which is defined by rewriting the Borel transform as
\be
\tB_b(u) = \fr{1}{(1-u)^{2s}} \sum_{n=0}^\infty \widetilde{B}_n^{(b,s)} u^n.
\ee
While this rewriting obviously has no effect on the full expression, at any finite order in the series, the parameter $s$ allows us to better model the behavior as $g\ra\infty$.

Given the first $N+1$ terms in the perturbative expansion of $F(g)$, we can therefore compute the truncated resummation
\be
F^{(N)}_B(g) = \fr{1}{g^{b+1}} \sum_{n=0}^N \widetilde{B}_n^{(b,s)} \int_0^1 du \fr{dt}{du} t^b(u) \, e^{-t(u)/g} \fr{u^n}{(1-u)^{2s}},
\label{eq:FinalTruncResum}
\ee
which approaches $F_B(g)$ as $N \ra \infty$. Note that while the exact function $F_B(g)$ is independent of both the Le Roy parameter $b$ and the summation variable $s$, at any finite truncation there is still some residual dependence on these two variables, which we can use to improve the accuracy of our results. Specifically, following \cite{Serone}, we choose the values of $b$ and $s$ to minimize the function
\be
\De F_B^{(N)} \equiv (\p_b F_B^{(N)})^2 + (\p_s F_B^{(N)})^2 + \Big(|F_B^{(N)} - F_B^{(N-1)}| - |F_B^{(N-1)} - F_B^{(N-2)}|\Big)^2.
\label{eq:MinFunc}
\ee
We can then obtain a rough estimate of the associated error by varying the parameters about the best-fit values $b_0,s_0$ and measuring the resulting shift in $F_B^{(N)}$.

\subsection{Borel Resumming Mass Gap}

Now that we have reviewed the general resummation procedure, let's apply it to our specific example of $\mb_\gap^2(\lb_\LC)$. To do so, we first need to construct the perturbative expansion of this function to some order in $\lb_\LC$. As discussed in section~\ref{sec:PertMatch}, we can do this by applying our perturbative map between couplings~\eqref{eq:PertMatching} to the expansion of $\mb_\gap^2(\lb_\ET)$, replacing each power of $\lb_\ET$ with a series in $\lb_\LC$ to obtain
\be
\begin{aligned}
\mb_\gap^2\big(\lb_\ET(\lb_\LC)\big) &= 1 - \fr{3}{2} \, \lb_\LC^2 + \fr{9}{\pi} \, \lb_\LC^3 - 11.4906 \, \lb_\LC^4 + 52.7576 \, \lb_\LC^5  \\
& \qquad - \, 287.357 \, \lb_\LC^6 + 1758.23 \, \lb_\LC^7 - 11901.4 \, \lb_\LC^8 + O(\lb_\LC^9).
\end{aligned}
\ee
In the language of the previous subsection, this expansion of the mass gap to $O(\lb_\LC^N)$ defines our truncated series $F^{(N)}(g)$. Using the terms in this sum, we can construct the Borel transform coefficients $\widetilde{B}^{(b,s)}_n$, then multiply these coefficients by the inverse Borel transform integrals given in eq.~\eqref{eq:FinalTruncResum} to obtain the resummed function $F^{(N)}_B(g)$, which depends explicitly on the two parameters $b$ and $s$.

For a given truncation level $N$, we then scan over values of $\lb_\LC$ between 0.2 and 0.9, and for each value of the coupling, determine the $b$ and $s$ which minimize the function $\De F^{(N)}_B$ given in eq.~\eqref{eq:MinFunc}. This gives us a distribution of values for $b$ and $s$, and we use the central values $b_0,s_0$ to define our final resummed function. For example, in the case of $N=8$, which is the highest truncation level we consider, we obtain the values
\be
\begin{aligned}
b^{(N=8)}_0 &= 4.14, \quad \De b^{(N=8)} = 0.14, \\
s^{(N=8)}_0 &= 2.84, \quad \De s^{(N=8)} = 0.0065.
\end{aligned}
\ee
where $\De b^{(N)},\De s^{(N)}$ simply correspond to the difference between the highest and lowest values of $b,s$ obtained by scanning over couplings.

\begin{figure}[t!]
\begin{center}
\includegraphics[width=0.6\textwidth]{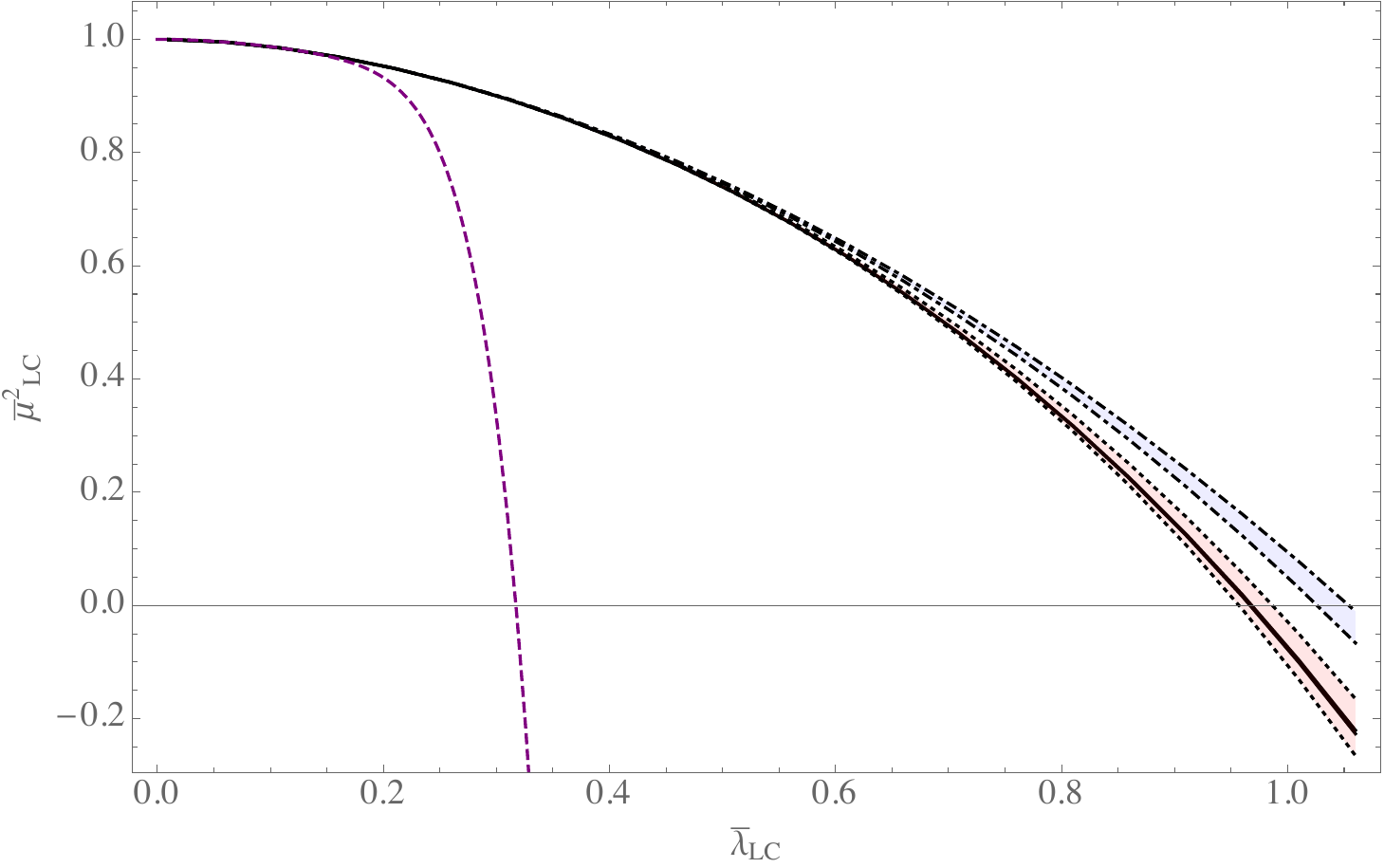}
\caption{Plots of estimates of the gap from 6th, 7th, and 8th order, for (blue, dot-dashed), (red, dotted), and (black, solid), respectively.  The upper and lower lines are the upper and lower values from moving $b,s$ away from their ``best-fit'' values as described in the text.
Additional errors due to the change from one order in perturbation theory to the next can be read off by comparing the different lines. We also show in purple, dashed, a plot of the Taylor series truncated at $\lambda^8$. }
\label{fig:BorelGap}
\end{center}
\end{figure}

Fig.~\ref{fig:BorelGap} shows the resulting resummed $\mb^2_\gap(\lb_\LC)$ for $N = 6$ (blue, dot-dashed), $7$ (red, dotted), and $8$ (black, solid). For each value of $N$, the upper and lower lines correspond to $b^{(N)}_0 \pm \De b, s^{(N)}_0 \pm \De s$, where $\De b, \De s$ correspond to the difference between the maximum and minimum values of $b,s$ obtained over \emph{all} considered values of the coupling $0.2 \leq \lb_\LC \leq 0.9$ and truncation order $6 \leq N \leq 8$,
\be
\De b = 0.58, \quad \De s = 0.56.
\ee

As we can see, there is a significant correction in going from $N=6$ to $N=7$, but by $N=8$ the sum appears to have largely converged for all $\lb_\LC$ below the critical coupling, where the mass gap closes. The estimated error for $N=8$ is much smaller than the previous orders, which indicates that this result is largely independent of $b$ and $s$, as we'd expect for the fully resummed function. However, it is worth pointing out that simply varying the resummation parameters clearly underestimates the overall error (at least for low $N$), since the error bars for $N=6$ do not contain the results for $N=7,8$.

\begin{figure}[t!]
\begin{center}
\includegraphics[width=0.32\textwidth]{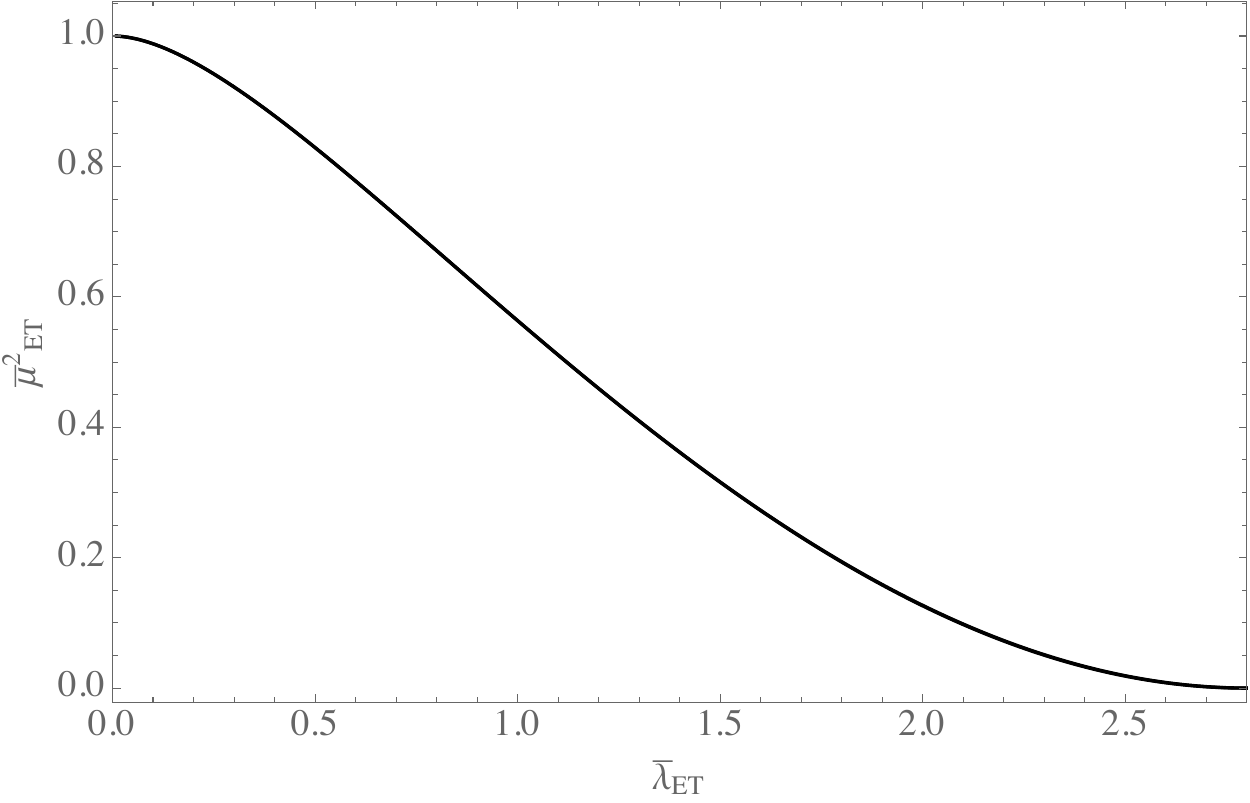}
\includegraphics[width=0.32\textwidth]{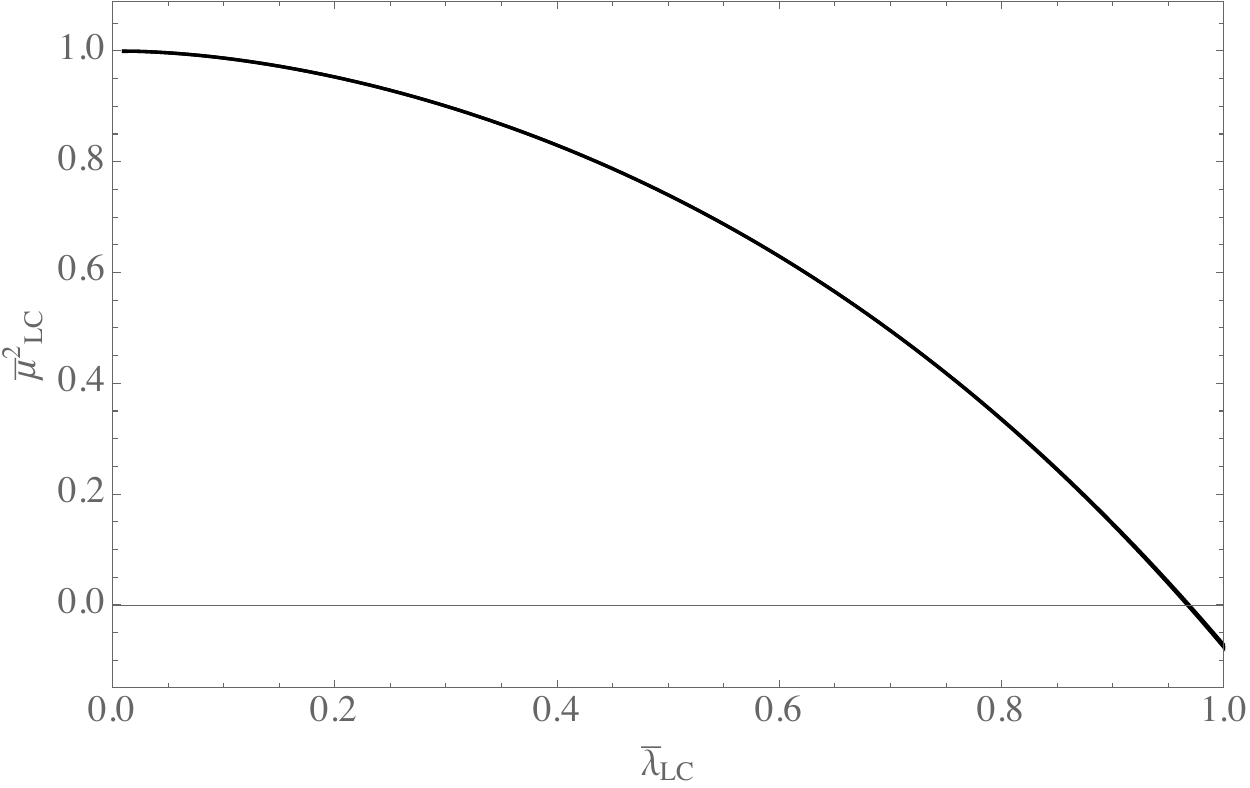}
\includegraphics[width=0.32\textwidth]{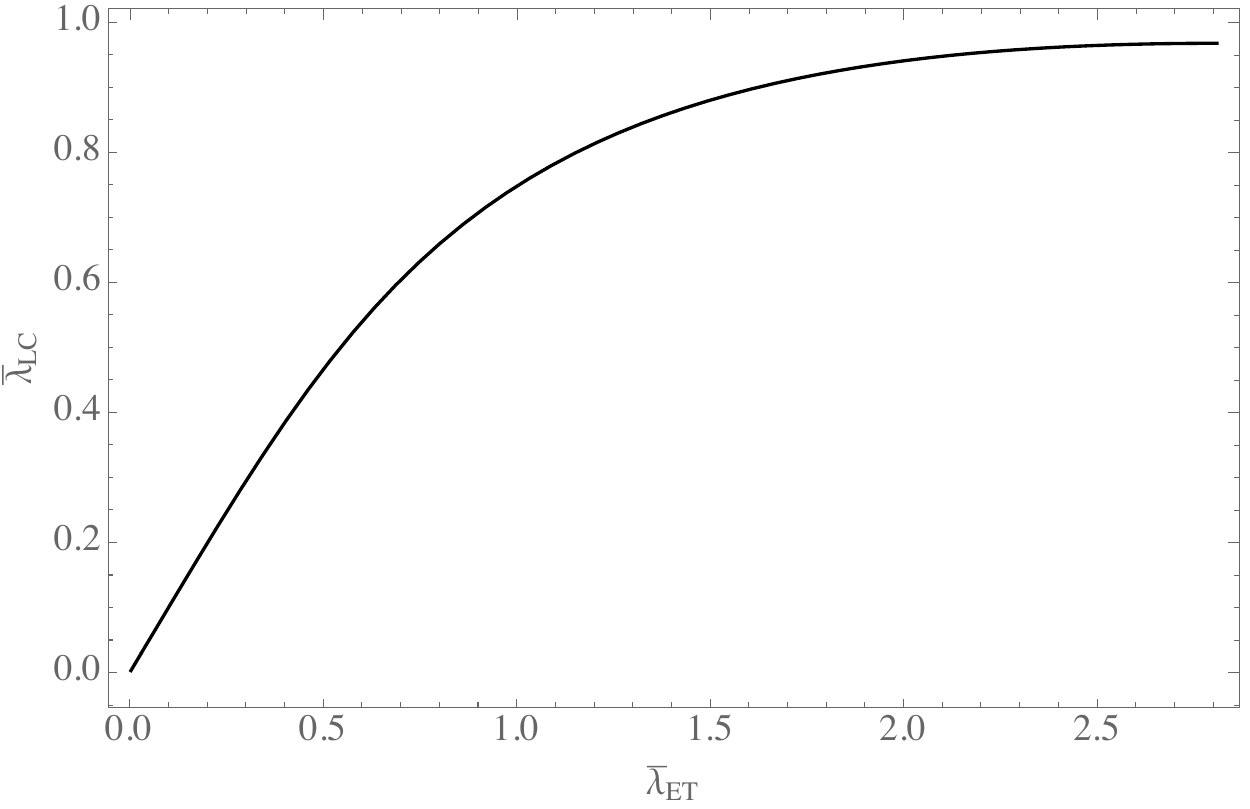}
\caption{{\it Left}: Gap $\bar{\mu}^2$ as a function of $\bar{\lambda}_{\rm ET}$, from Borel resumming its perturbation series at eighth order. {\it Center}: Gap $\bar{\mu}^2$ as a function of $\bar{\lambda}_{\rm LC}$ from Borel resumming its perturbation series, also at eighth. {\it Right}: Inferred map $\bar{\lambda}_{\rm LC}(\bar{\lambda}_{\rm ET})$ from imposing $\bar{\mu}^2_{\rm LC}(\bar{\lambda}_{\rm LC})/\bar{\lambda}_{\rm LC}=\bar{\mu}^2_{\rm ET}(\bar{\lambda}_{\rm ET})/\bar{\lambda}_{\rm ET}$. In the left and center plot, errors (barely visible) are calculated as in Fig.~\ref{fig:BorelGap}. 
}
\label{fig:Map}
\end{center}
\end{figure}

Using only ET perturbation theory data, combined with the perturbative map between ET and LC couplings in eq.~\eqref{eq:PertMatching}, we've therefore constructed a numerically resummed approximation to the \emph{nonperturbative} LC mass gap $\mb_\gap^2(\lb_\LC)$. Similarly, we can use this same technique to Borel resum $\mb_\gap^2(\lb_\ET)$ (which simply reproduces the results of \cite{Serone}). These two results, both at truncation level $N=8$, are shown in the center and left of Fig.~\ref{fig:Map}, respectively.

We can now use these two intermediate functions to construct the nonperturbative map $\lb_\LC(\lb_\ET)$. Specifically, we can identify which points in the two plots correspond to the same physical theory, parameterized by the ratio $\mb_\gap^2 / \lb$. In other words, for each value of $\lb_\ET$, we can use the left plot of Fig.~\ref{fig:Map} to determine the corresponding value of $\mb_\gap^2 / \lb_\ET$. We can then use the middle plot to determine which value of $\lb_\LC$ has the \emph{same} $\mb_\gap^2 / \lb_\LC$, thus giving us a map between bare couplings in ET and LC quantization, shown in the right plot of Fig.~\ref{fig:Map}.  In particular, we find the following different values for the coupling $\bar{\lambda}_*$ at the critical point in the two quantizations:
\be
\bar{\lambda}_{\rm ET *} = 2.81 , \qquad \bar{\lambda}_{\rm LC *} = 0.97,
\ee
where the former is just reproducing the calculation in \cite{Serone}. These values can be compared with previous results obtained using renormalized Hamiltonian truncation~\cite{Elias-Miro:2017xxf,Elias-Miro:2017tup}, lattice Monte Carlo methods~\cite{Bronzin:2018tqz}, matrix product states~\cite{Milsted:2013rxa}, and the tensor network renormalization group~\cite{Kadoh:2018tis} for the case of ET quantization, and symmetric polynomials~\cite{Burkardt:2016ffk}, discrete lightcone quantization~\cite{Harindranath:1988zt}, and lightcone conformal truncation~\cite{Anand:2017yij} for the case of LC quantization, listed in table~\ref{fig:CriticalCouplings}.\footnote{See also \cite{Romatschke:2019rjk} for a recent new method for calculating the mass gap and vacuum energy.}

{\setlength{\extrarowheight}{2pt}%
\begin{table}[t!]
\begin{center}
\begin{tabular}{|lc|lc|}
\hline
ET Method & $\bar{\lambda}_{\ET*}$ & LC Method & $\bar{\lambda}_{\LC*}$ \\ \hline
Borel~\cite{Serone} & $2.807 \pm .034$ & Borel (this work) & $0.97 \pm .01$ \\
HT~\cite{Elias-Miro:2017xxf,Elias-Miro:2017tup} & $2.76 \pm .03$ & LCT~\cite{Anand:2017yij} & $0.96 \pm .02$ \\
Lattice MC~\cite{Bronzin:2018tqz} & $2.764 \pm .004$ & Symm.\ Poly.~\cite{Burkardt:2016ffk} & $1.10 \pm .03$ \\
MPS 1~\cite{Milsted:2013rxa} & $2.769 \pm .002$ & DLCQ~\cite{Harindranath:1988zt} & $1.4$ \\
MPS 2~\cite{Milsted:2013rxa} & $2.7625 \pm .0008$ & & \\
TRG~\cite{Kadoh:2018tis} & $2.728 \pm .014$ & & \\ \hline
\end{tabular}
\end{center}
\caption{Computed values of the critical coupling $\bar{\lambda}_*$ from various methods in both ET (left) and LC (right) quantization.}
\label{fig:CriticalCouplings}
\end{table}

\subsection{Linear Closing of the Gap in LC}

Before testing our nonperturbative map $\lb_\LC(\lb_\ET)$ in section~\ref{sec:Tests}, let's first briefly mention an important subtlety in our two Borel resummed functions $\mb_\gap^2(\lb_\ET)$ and $\mb_\gap^2(\lb_\LC)$. As discussed in~\cite{Serone}, the accuracy of this truncated resummation procedure at $N=8$ is still quite sensitive to which power $\mb_\gap^\alpha$ we choose to Borel resum. In particular, our choice of $\alpha$ determines how the Borel resummed mass gap vanishes as we approach the critical point. We can easily understand this by noting in eq.~\eqref{eq:FinalTruncResum} that $F^{(N)}(g)$ is analytic in $g$, which means that as we approach a critical point, generically the resummed function will vanish linearly with $g$. If we choose to Borel resum $\mb_\gap^\alpha(\lb)$, we thus expect the inferred mass gap to vanish as
\be
\mb_\gap(\lb) \sim  |\lb - \lb_*|^{\fr{1}{\alpha}}, \quad \lb \ra \lb_* \quad (\textrm{Truncated Borel}).
\ee
However, we know that the behavior of the \emph{exact} mass gap function is set by the critical exponent $\nu$ associated with the lowest singlet operator in the IR fixed point,
\be
\mb_\gap(\lb) \sim  |\lb - \lb_*|^\nu, \quad \lb \ra \lb_* \quad (\textrm{Exact}).
\ee
The optimal choice of $\alpha$ for convergence of the truncated resummation is thus $\alpha = \fr{1}{\nu}$.\footnote{In principle, we can choose any value of $\alpha$, and the resulting mass gap will converge to the exact answer as we take the truncation level $N \ra \infty$. We are simply noting that, in practice, we can improve the rate of convergence near the critical point if we include knowledge of the critical exponent $\nu$ in the Borel resummed function.}

For this particular example of $\phi^4$ theory in $d=2$, the critical point is in the same universality class as the 2D Ising model, with the known critical exponent $\nu = 1$. In constructing the left plot of Fig.~\ref{fig:Map}, we therefore technically chose to resum the function $\mb_\gap(\lb_\ET)$, as was done in~\cite{Serone}, then squared the result to obtain $\mb_\gap^2$.

However, there is a further subtlety in Borel resumming the LC mass gap, which is that \emph{the mass gap does not close as $|\lb - \lb_*|^\nu$ in LC quantization}. To understand this, we can study the mass gap from a Hamiltonian perspective. At linear order around the critical point,  the LC Hamiltonian can be written in the form
\be
P_+(\lambda) = P_{+*}(\lambda_*) + \fr{1}{2\lb_*}(\lambda_*-\lambda) \int dx \, \phi^2(x).
\ee
Moreover, the Hamiltonian  $P_+$ is proportional to $\mu^2$,
\be
\mu^2 = 2P_+P_-.
\ee
We therefore can write the LC mass gap squared as 
\be
\mu^2_{\gap,\LC} = \fr{1}{2\lb_*} |\lambda - \lambda_*|  \<1|\phi^2(0)|1\>.
\ee
In order for the LC mass gap to vanish as $|\lb-\lb_*|^\nu$, the expectation value of $\phi^2$ in the first excited state would therefore need to vanish as $|\lb-\lb_*|^{2\nu-1}$. However, this expectation value is necessarily \emph{positive} at the critical point. One simple way to see this is to expand the first excited state in terms of free mass eigenstates $|m_i^2\>$ (i.e.~$|1\> = \sum_i c_i |m_i^2\>$), such that the expectation value $\< 1 | \phi^2 |1\>$ is clearly a sum of positive terms:
\be
 \<1|\phi^2(0)|1\> =  \sum_i |c_i|^2 \frac{m_i^2}{m^2}.
\ee
We therefore see that in $\phi^4$ theory $\mu^2_{\gap,\LC}$ must vanish \emph{linearly} with $\lb$, regardless of the critical exponent $\nu$.\footnote{Technically, $\mu^2_{\gap,\LC}$ could vanish as a smaller than linear power, if the expectation value of $\phi^2$ diverged as $\lb \ra \lb_*$. However, based on the Borel resummation results in this work and separate LC Hamiltonian truncation results~\cite{Elliott:2014fsa,Burkardt:2016ffk,Chabysheva:2016ehd,Anand:2017yij}, this expectation value appears to remain finite in $d=2$, such that $\mu^2_{\gap,\LC}$ vanishes linearly.} Based on this observation, we can optimize the convergence of our truncated sum by Borel resumming the function $\mb_\gap^2(\lb_\LC)$, which was done to construct the middle plot of Fig.~\ref{fig:Map}. For contrast, we have also shown the results in Fig.~\ref{fig:GapFromBorelMass} if we instead resum $\mb_\gap(\lb_\LC)$; as one can see, the convergence rate is clearly worse than in Fig.~\ref{fig:BorelGap}.

\begin{figure}[t!]
\begin{center}
\includegraphics[width=0.48\textwidth]{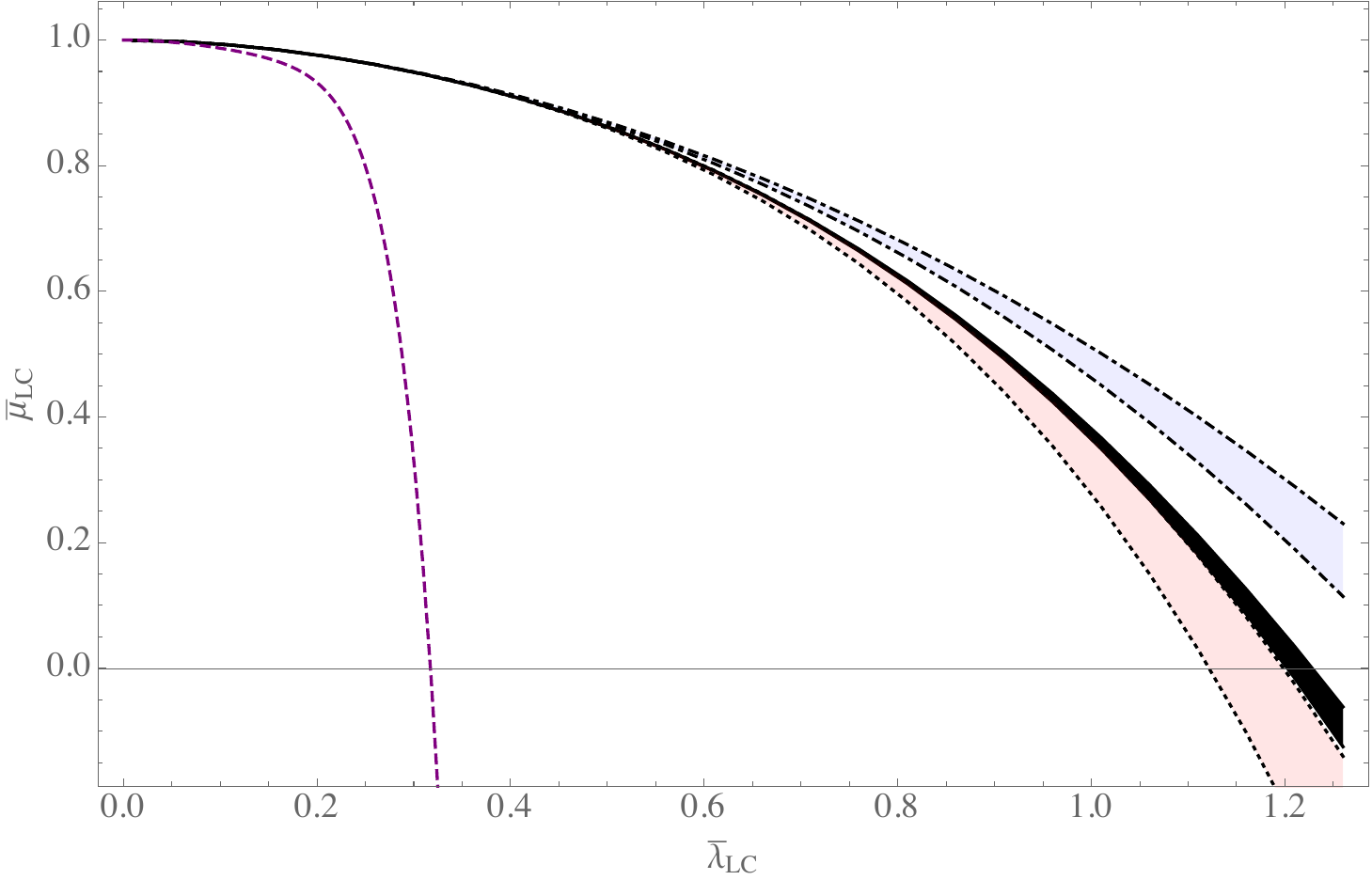}
\includegraphics[width=0.48\textwidth]{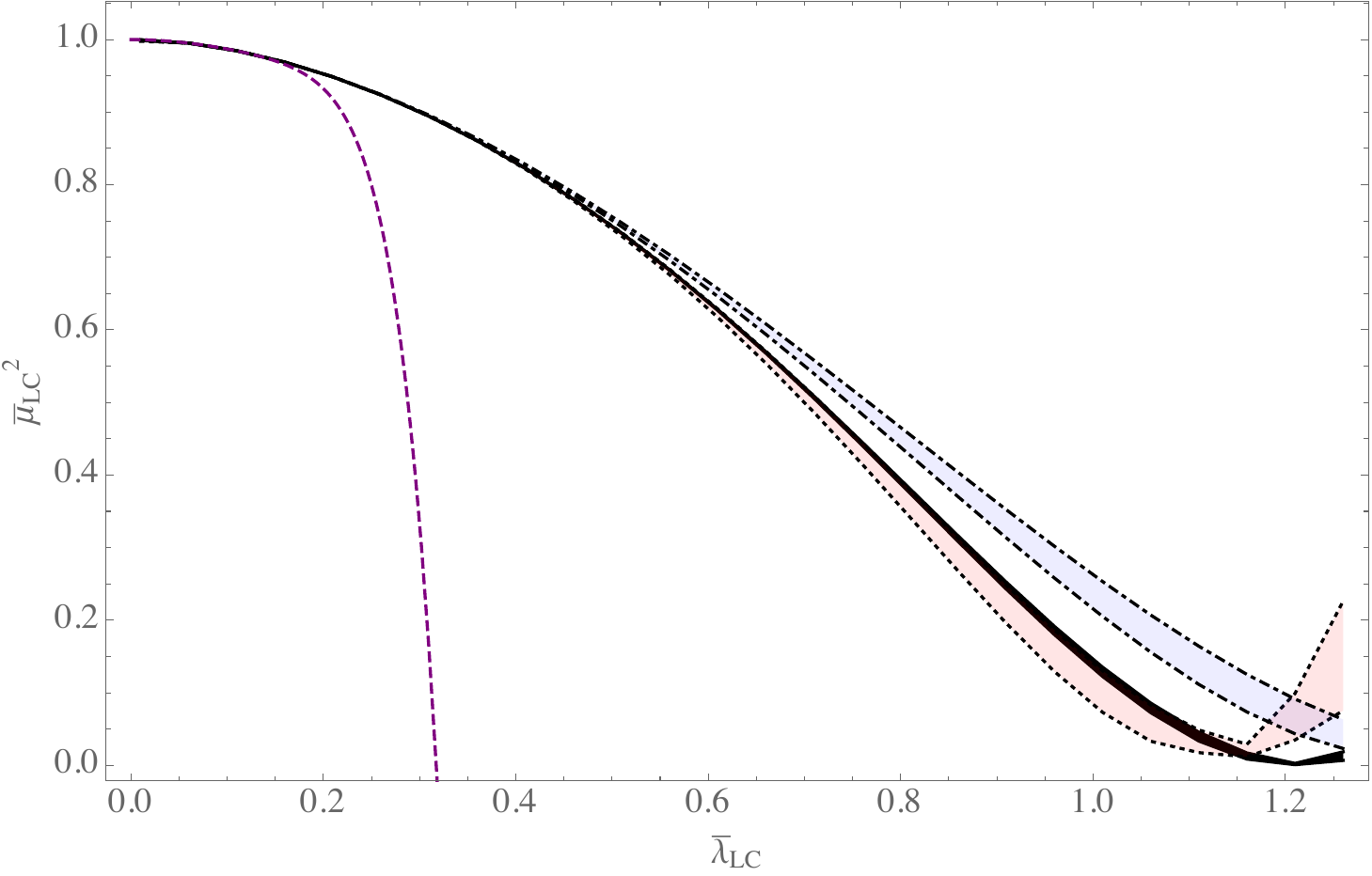}
\caption{Same as Fig.~\ref{fig:BorelGap}, but for Borel resumming $\bar{\mu}_{\rm LC}$ rather than Borel resumming $\bar{\mu}^2_{\rm LC}$ in LC. The right plot just shows the square of the left plot.}
\label{fig:GapFromBorelMass}
\end{center}
\end{figure}

More generally, in any theory where the coefficient of a relevant UV operator $\Ocal_R$ can be tuned to reach a IR fixed point, we can look at the expectation value of that operator in the first excited state to determine how the LC mass gap behaves near the critical point. If this expectation value is positive and finite, then $\mu_{\gap,\LC}^2$ will close linearly in the associated coupling.

One interesting consequence of this observation is that in such theories the map from ET to LC couplings must account for this differing critical behavior. Specifically, we expect the scaling relation
\be
|\lb_\LC - \lb_{\LC,*}| \sim |\lb_\ET - \lb_{\ET,*}|^{2\nu} \qquad (\lb \ra \lb_*),
\ee
as demonstrated in the right plot of Fig.~\ref{fig:Map}, where our inferred map $\lb_\LC(\lb_\ET)$ approaches the critical point quadratically in the ET coupling.

In other words, \emph{the map between ET and LC couplings contains information about critical exponents}. In principle, if one could construct this map directly, rather than from matching ET and LC results, then one would have a method of directly computing the critical exponent $\nu$. Though we currently have no method for doing so for this particular model, in appendix~\ref{app:ONModel} we consider the 3D $O(N)$ model at large $N$, where we \emph{can} directly calculate the map $\lb_\LC(\lb_\ET)$. In this example, we can explicitly see how the linear critical behavior in LC quantization is corrected by the map to reproduce the appropriate critical exponent in ET quantization.

\section{ Tests of the Mapping}
\label{sec:Tests}

In the previous section, we extracted a map between the LC coupling $\bar{\lambda}_{\rm LC}$ and ET coupling $\bar{\lambda}_{\rm ET}$ from the ET perturbative expansion of $\mu_\gap^2$ by assuming that the mass gap was Borel resummable in both quantization schemes.  In this section, we will look at some tests of this map by using it to compare physical quantities that have been computed by Hamiltonian truncation methods in both quantizations. The LC data was initially computed in~\cite{Anand:2017yij}, using the method of LC conformal truncation. For these results, we used a basis of primary operators in massless scalar field theory up to $\Dmax=33$ (with 5084 states in the $\mathbb{Z}_2$-odd sector) to then extrapolate $\Dmax \ra \infty$. The ET data was initially computed in~\cite{Rychkov:2014eea} using a basis of free massive energy eigenstates on $S^1$ with radius $L=10$ (in units of the bare mass $m$) and energy cutoff $E_{\max} = 20$ (12801 odd sector states).\footnote{We thank Lorenzo Vitale for kindly providing us with these ET results.}

\subsection{Mass Gap}

For our first check of the map $\bar{\lambda}_{\rm ET}(\bar{\lambda}_{\rm LC})$, we consider the mass gap $\mu_{\rm gap}^2$ computed by Hamiltonian truncation in both quantization schemes.  Of course, the idea of the previous section was that one should be able to obtain the gap in either quantization by Borel-resumming its perturbation series. With Hamiltonian truncation, we can check this proposal directly, by computing the gap numerically and using the map obtained from Borel resummation to compare the two quantization schemes.

First, in the left plot of Fig.~\ref{fig:Comparison}, we show the Hamiltonian truncation result for $\bar{\mu}_\gap^2$ computed in both ET (black, solid) and LC (red, dashed).  More precisely, in either quantization we can vary the bare mass-squared $m^2$ and the bare coupling $\lambda$, compute $\bar{\mu}_\gap^2 \equiv \mu^2_{\rm gap}/m^2$, and plot the result as a function of $\bar{\lambda} \equiv \lambda/m^2$.  At very small couplings $\bar{\lambda} \lesssim 0.1$, the two curves are very similar, but quickly diverge at larger couplings where it becomes crucial to take into account the fact that the bare parameter $\bar{\lambda}$ in the two quantizations does not match.

\begin{figure}[t!]
\begin{center}
\includegraphics[width=0.42\textwidth]{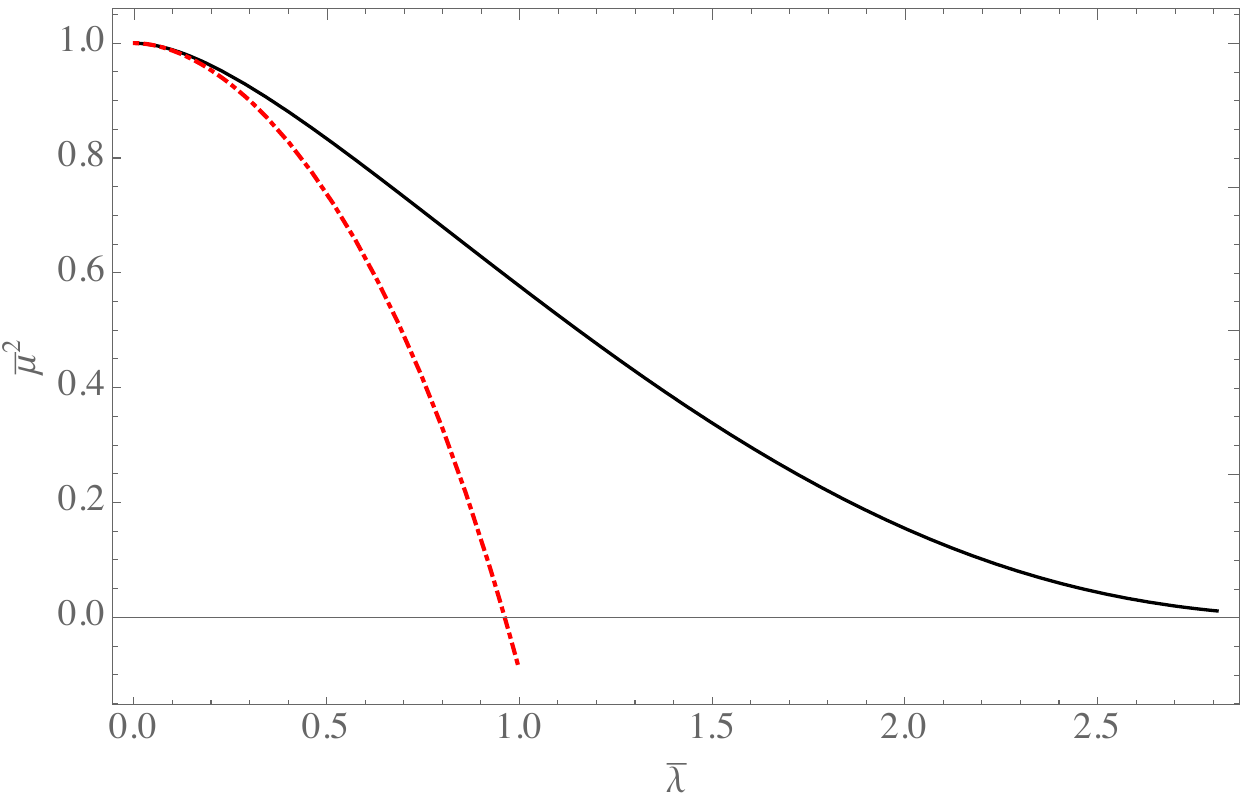}
\includegraphics[width=0.42\textwidth]{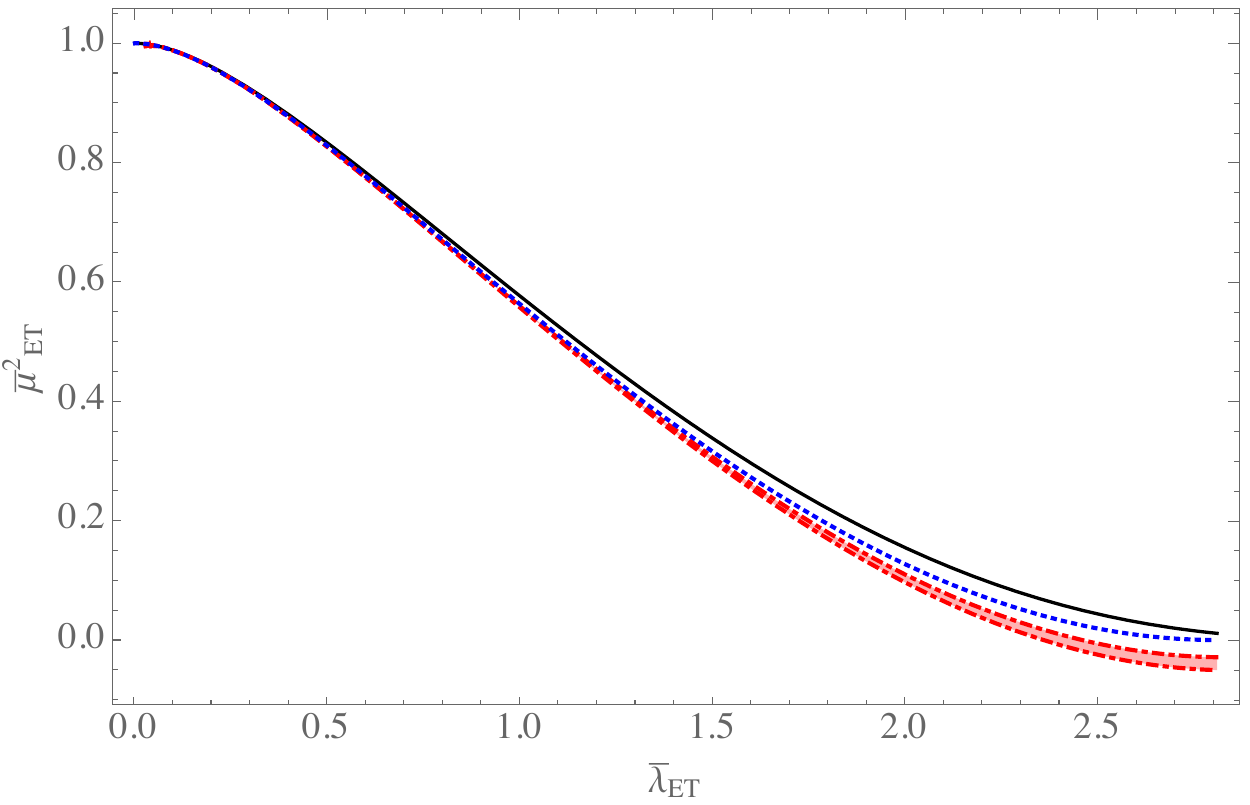}
\caption{{\it Left}: Comparison of  $\bar{\mu}^2= \frac{\mu_{\rm gap}^2}{m^2}$ as a function of $\bar{\lambda}= \frac{\lambda}{m^2}$, for ET (black, solid) and LC (red, dashed) quantization from numerical conformal truncation analysis.  The mismatch is apparent, due to the fact that the bare parameter $m$ should not be the same in both quantization schemes.  {\it Right}: Comparison of  $\bar{\mu}^2_{\rm ET} = \frac{\mu_{\rm gap}^2}{m_{\rm ET}^2}$ as a function of $\bar{\lambda}_{\rm ET}$ after applying the map from $\lambda_{\rm ET}$ to $\lambda_{\rm LC}$ in Fig.~\ref{fig:Map}.  The black, solid line is the result from an ET numerical conformal truncation analysis, the same as in the left plot.  The red, dashed line is the result from a LC numerical conformal truncation analysis, after applying the map.  The blue, dotted line is the gap in ET quantization from Borel resumming the ET perturbation series. 
}
\label{fig:Comparison}
\end{center}
\end{figure}

In the right plot of Fig.~\ref{fig:Comparison}, we have used the map $\bar{\lambda}_{\rm LC} \leftrightarrow \bar{\lambda}_{\rm ET}$ derived in the previous section using Borel resummation to correct the mismatch.  The ET value of $\bar{\mu}_\gap^2$ as a function of $\bar{\lambda}_\ET$ is the same as in the left plot.  The LC result $\bar{\mu}^2_{\rm LC}(\bar{\lambda}_{\rm LC})$ from the left plot has been converted to ET by substituting it into the formula
\be
\bar{\mu}^2_{\rm ET}(\bar{\lambda}_{\rm ET}) = \frac{\bar{\lambda}_{\rm ET}}{\bar{\lambda}_{\rm LC}(\bar{\lambda}_{\rm ET})} \, \bar{\mu}^2_{\rm LC}\Big(\bar{\lambda}_{\rm LC}(\bar{\lambda}_{\rm ET})\Big).
\ee
In effect, we have used the gap computed using Borel resummation to ``undo'' the difference in the gap computed using conformal truncation, such that the converted LC truncation results now match those of ET. We also show some spread in this LC result, coming from the spread in the 8th order result in Fig.~\ref{fig:BorelGap}.  Finally, we also show for comparison the ET result (blue, dashed) obtained from Borel resumming its perturbation series (from the left plot in Fig.~\ref{fig:Map}).

\subsection{Residue at Single-Particle Pole}

For our second check of the map between bare parameters, we compute the residue $Z$ of the single-particle pole in the scalar two point function:
\be
G(p) = \frac{i Z}{p^2-\mu_\gap^2+ i \epsilon} + \dots .
\ee
Equivalently, $Z$ is defined in terms of the matrix element of the field $\phi$ between the ground state and the first excited state,
\be
Z = |\< \Omega | \phi(0) |1\>|^2.
\ee

\begin{figure}[t!]
\begin{center}
\includegraphics[width=0.7\textwidth]{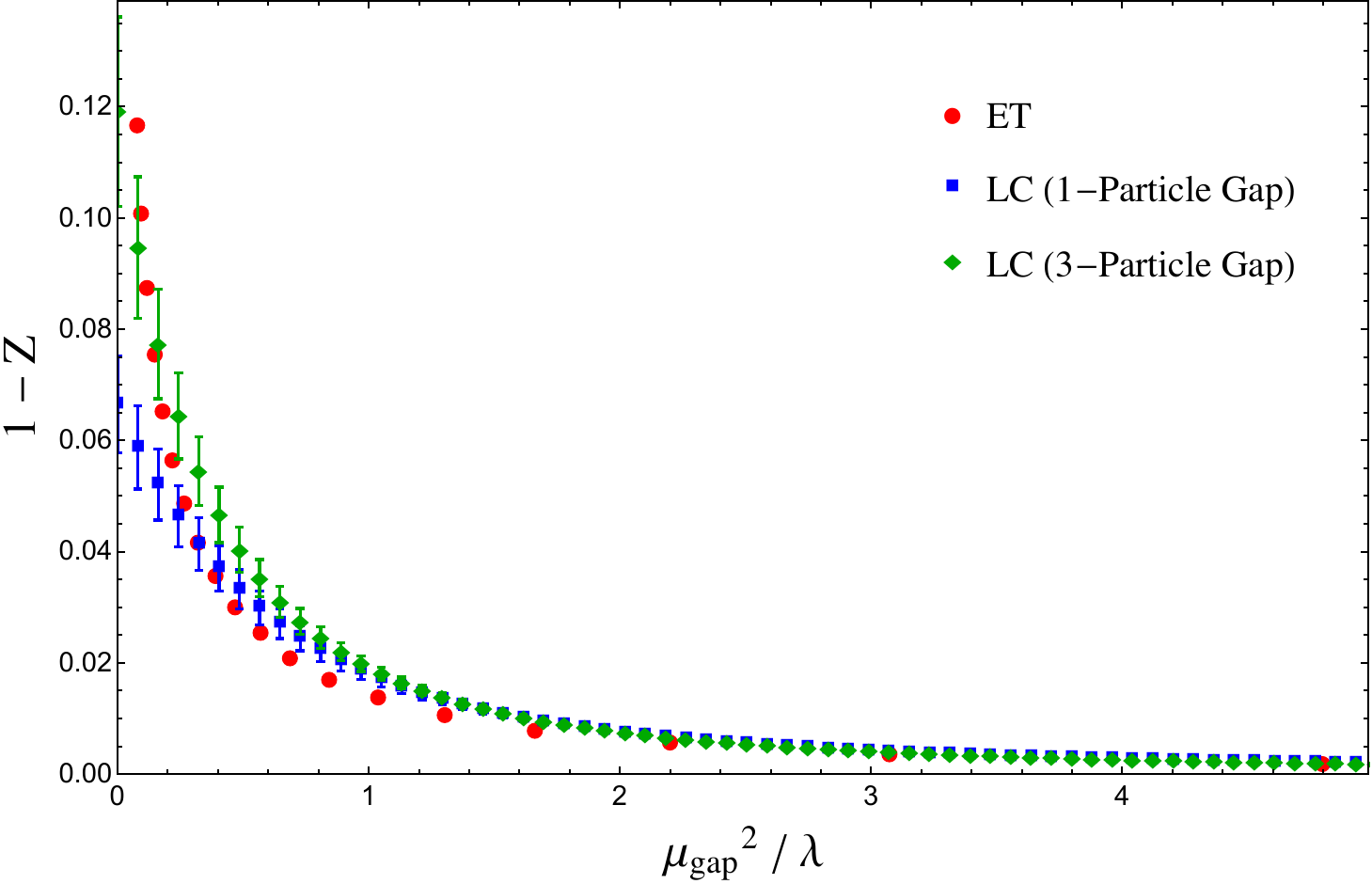}
\caption{Comparison of $1-Z$ as a function of $\mu_\gap^2 / \lambda$, in both ET (red) and LC (blue, green). The two LC results were extrapolated from $\Dmax=33$ data using both the 1-particle (blue) and 3-particle (green) thresholds to define the mass gap. Their disagreement near the critical point indicates that the LC truncation data has not fully converged for $\mu_\gap^2 / \lambda \lesssim 0.3$.}
\label{fig:Zcomparison}
\end{center}
\end{figure}

By definition, $Z = 1$ in the free theory (at $\lb=0$), but decreases from 1 at finite coupling. Fig.~\ref{fig:Zcomparison} shows the numerically computed deviation of $Z$ from the free value in both ET and LC quantization, as a function of the dimensionless ratio $\mu_\gap^2 / \lambda$. In constructing our nonperturbative map, we identified ET and LC theories with the same value for this ratio, which means the two schemes should yield the same result for the physical observable $Z$.

For the LC results, we have used two different definitions of the mass gap. The first (in blue), was obtained using the single-particle mass, $\mu_\gap^2 \equiv \mu_1^2$, while the other (green) was obtained by using the three-particle threshold, $\mu_\gap^2 \equiv \fr{1}{9} \mu_3^2$. In principle, these two definitions should be equivalent, but at any finite truncation $\mu_3 > 3\mu_1$. The comparison between these two extrapolated plots thus provides us with an indication of the error, along with the estimated error bars obtained by varying the slope of the extrapolation. As we can see, the two LC results are in good agreement until $\mu_\gap^2 / \lambda \lesssim 0.3$, which indicates that the results at $\Dmax=33$ have not fully converged near the critical point. This disagreement near $\mu_\gap =0$ is largely due to the fact that the single-particle eigenvalue necessarily reaches zero \emph{before} the three-particle threshold for any truncation.

We can then compare these two LC results to the ET data (red), which was calculated using the single-particle mass to define the gap. As we can see, all three results are consistent until close to the critical point. This agreement in the observable $Z$ indicates that at a fixed value of $\mu_\gap^2/\lambda$ both ET and LC truncation are describing the same theory, confirming our procedure for matching bare parameters.\footnote{In this case, we have not literally used our mapping of bare parameters, but rather went directly to $Z$ as a function of $\mu^2_{\rm gap}/\lambda$ in both quantizations. We have already seen that our map matches $\mu^2_{\rm gap}/\lambda$ in terms of the bare parameters to reasonably high accuracy, so in principle there is not much difference between first writing $Z$ in terms of $\bar{\lambda}_{\rm ET}$ and $\bar{\lambda}_{\rm LC}$ and then mapping, versus directly expressing $Z$ in terms of $\mu_{\rm gap}^2/\lambda$ in each quantization.  In practice, we have  found that extrapolating to infinite $\Delta_{\rm max}$ typically is more accurate when physical quantities are expressed in terms of other physical quantities, rather than in terms of bare parameters.}

\section{Future Directions}
\label{sec:future}

The main goal of this paper has been to obtain a deeper understanding of the relation between ET and LC quantization, focusing on the special case of $\lambda \phi^4$ theory in 2d.  In principle, the same analysis we have applied here could be done for $\lambda \phi^4$ in 3d.  The main challenges in 3d compared to 2d are due to the fact that the calculations all become computationally more expensive.  The perturbative analysis of \cite{Serone} in 3d would require doing loop integrals with a larger phase space and additional UV divergences, making it more challenging to go to $\CO(\lb^8)$.  The tests from comparing to conformal truncation results also become more difficult mainly due to the larger number of states at each level in higher dimensions.  Nevertheless, perhaps with available resources these obstacles could be overcome.

More generally, it is important to understand when the effect of LC zero modes is to just shift the bare parameters of the theory.  While obtaining a detailed map between the bare parameters is likely an impractically difficult task in most cases, one could hope to prove that such a map exists provided that certain simple criteria are satisfied. For most practical purposes, such a proof would be as good as the map itself, since usually one is interested in the relation between different physical quantities in the theory rather then their (usually scheme-dependent) dependence on the bare parameters.  A natural conjecture in the spirit of the analysis of this paper is that any time the perturbative effect of zero modes can be absorbed into a shift in the bare parameters, their nonperturbative effect can be as well.  A closely related question is whether or not Borel resummability of a physical quantity in ET perturbation theory implies its Borel resummability in LC.  We have essentially assumed that this is true of 2d $\lambda \phi^4$ theory in this paper, and have attempted to test this assumption numerically, but a proof would of course be more desirable.

There have been many previous studies of 2d $\lambda \phi^4$ theory using lattice MC methods~\cite{Bronzin:2018tqz,Bosetti:2015lsa,Schaich:2009jk}, tensor networks~\cite{Milsted:2013rxa,Kadoh:2018tis}, and the density matrix renormalization group~\cite{Sugihara:2004qr}. However, those studies have largely focused on the critical point, in order to extract the critical coupling $\bar{\lambda}_*$, as well as critical exponents. It would be very useful if such methods could be used to extract the $\bar{\lambda}$-dependence of observables such as the mass gap, vacuum energy, and vev $\<\phi^2\>$ away from the critical point. Such data would provide a useful additional check of the map between ET and LC quantization, as well as insight into the nonperturbative structure of $\lambda \phi^4$ theory, more generally.

Finally, we have focused our analysis on the symmetry preserving phase, $\< \phi \>=0$, but it would be very interesting to understand the symmetry-broken phase of the theory as well.  In this case, because of the apparent triviality of the vacuum in LC quantization, we expect that one has to start with the Lagrangian expanded around the true vacuum.  As a result, the Lagrangian would have a $\phi^3$ term in addition to the quadratic and quartic.  A puzzle in this approach is that the coefficient of the $\phi^3$ interaction should not really be an independent parameter of the theory, which is fully determined in the original manifestly $\mathbb{Z}_2$-symmetric Lagrangian by only two parameters.  An additional constraint is provided by the fact that for the correct value of the $\phi^3$ coefficient as a function of the $\phi^2$ and $\phi^4$ coefficients, the spectrum of the theory must be invariant under $\phi^3 \rightarrow - \phi^3$.  In the case of the $O(N)$ 2d model, or more generally for theories with spontaneous breaking of continuous symmetries, one could also constrain the parameters of the theory by demanding that the spectrum contain  massless Goldstone bosons.  Potentially, such constraints could be used to fix the coefficient of $\phi^3$.  We leave these questions to future work.

\section*{Acknowledgments}

First we would like to thank Lorenzo Vitale for collaboration in the early stages of this project, for numerous discussions, and for graciously providing his results for the mass gap and $Z$ residue from ET Hamiltonian truncation.  We would also like to thank Marco Serone in particular, as well as 
Nikhil Anand, Jared Kaplan, and Zuhair Khandker, for valuable discussions. ALF and EK were supported in part by the US Department of Energy Office of Science under Award Number DE-SC0015845, and ALF in part by a Sloan Foundation fellowship.    ALF, EK and MW were also supported in part by the Simons Collaboration Grant on the Non-Perturbative Bootstrap. 

\appendix

\section{0d Example}
\label{sec:0dexample}

In this paper, we have mostly focused on the prescription in $\lambda \phi^4$ theory that assigns an effective value of $\bar{\lambda}_{\rm LC}(\bar{\lambda}_{\rm ET})$ in LC quantization corresponding to an ET computation according to
\be
m_{\rm LC}^2 = m_{\rm ET}^2 + 12 \lambda_{\rm ET} \< \phi^2\>_{\rm ET}.
\label{eq:mBurkardt2}
\ee
The theory is determined by the dimensionless combination $\bar{\lambda} = \frac{\lambda}{m^2}$. A problem with (\ref{eq:mBurkardt2}) is that applying it literally assigns the same lightcone $\bar{\lambda}_{\rm LC}$ to two different values of $\bar{\lambda}_{\rm ET}$.  In this appendix, we will analyze the analogous phenomenon in a lower dimensional example.

We will find it somewhat conceptually simpler to work in units with $\lambda=1$ in both quantizations.
The first point we want to make here is that we can state the puzzle of why (\ref{eq:mBurkardt2}) fails without referring to LC quantization.  Because the plant diagrams have no momentum dependence, their contributions can be exactly absorbed into a mass counterterm.  This is essentially the same argument that one uses when one normal-orders the action: one can simply drop the one-loop contribution to the two-point function since it can be absorbed into a (log divergent) shift in the mass term.  The next plant diagram occurs at three loops,
\be
\includegraphics[width=0.13\textwidth]{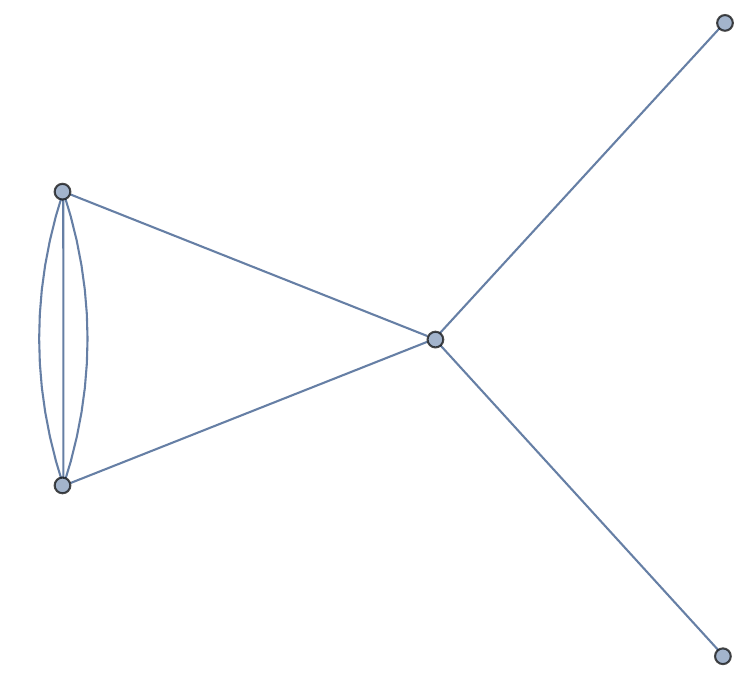} \begin{array}{c} \cong \left( m^2 \rightarrow m^2 + \frac{1}{m^4} \frac{63 \zeta(3)}{2 \pi^3}\right) \\ \\ \\  \end{array} .
\ee
This diagram can also be removed precisely with a counterterm.  We can continue in this way to any order in perturbation theory, without ever having to mention LC quantization or making any conjectures about what LC does.  In other words, in perturbation theory, we really can just remove all plant diagrams by defining a new mass term.

In this language, we can reinterpret the lower left plot in Fig \ref{fig:NonPertBurkFail} as a statement of what new mass $m_{\rm LC}^2$ we get as a function of the original mass $m_{\rm ET}^2$. At large $m_{\rm ET}^2$, the shift from $\< \phi^2\>$ is small compared to $m_{\rm ET}^2$, and so everything looks fine. However, as we decrease $m_{\rm ET}$, the new mass $m^2_{\rm LC}(m^2_{\rm ET})$ in the new effective ET description (where we have deleted all the plant diagrams)  receives larger and larger contributions from the plant diagrams, and eventually these more than compensate for the decreasing ``bare'' mass.  Even without doing any additional computations, we see that this procedure naively makes a prediction {\it for the results in ET quantization}, namely that two different values of $m^2_{\rm ET}$ should give the same physical results.  The problem is that this prediction is wrong.   
Stated this way, the failure of the prediction may seem surprising.  The plant diagrams really are present in the original ET computation, and they really do give a contribution to the mass, so it seems like they really should push the effective mass back up.

We can get more intuition about what is going wrong by asking this question in the following simpler toy ``model'':
\be
Z(m^2) \equiv \int_{-\infty}^\infty dx e^{-m^2 x^2 - x^4}  = \frac{1}{2} e^{\frac{m^4}{8}} m K_{\frac{1}{4}}\left(\frac{m^4}{8}\right),
\label{eq:0dInt}
\ee
where $K_\nu$ is a Bessel function.  A ``diagrammatic'' evaluation of this integral is just its series expansion in $1/m^2$:
\be
Z(m^2) = \frac{\sqrt{\pi}}{m} \left( 1 - \frac{3}{4 m^4} + \frac{105}{32 m^8} + \dots\right) =
 \frac{\sqrt{\pi}}{m} \sum_{n=0}^\infty \frac{1}{(-4m^4)^n}\frac{ (4 n-1)\text{!!}}{(n)!}.
 \label{eq:Zexp}
 \ee
One can easily compute the ``vev'' exactly
\be
\<x^2\> = \frac{1}{4} m^2
   \left(\frac{K_{\frac{3}{4}}\left(\frac{m^4}{8}\right)}{K_{\frac{1}{4}}\left(\frac{m^
   4}{8}\right)}-1\right) .
   \label{eq:x2vev0d}
   \ee
As we did in the 2d theory, we can plot an effective ``LC'' mass-squared as a function of the ``ET'' mass-squared:
\be
m_{\rm LC}^2 = m^2 + 6 \< x^2\> ,
\label{eq:Burkardt0d}
\ee
shown in Fig.~\ref{fig:Burkardt0d}. 

\begin{figure}[t!]
\begin{center}
 \includegraphics[width=0.4\textwidth]{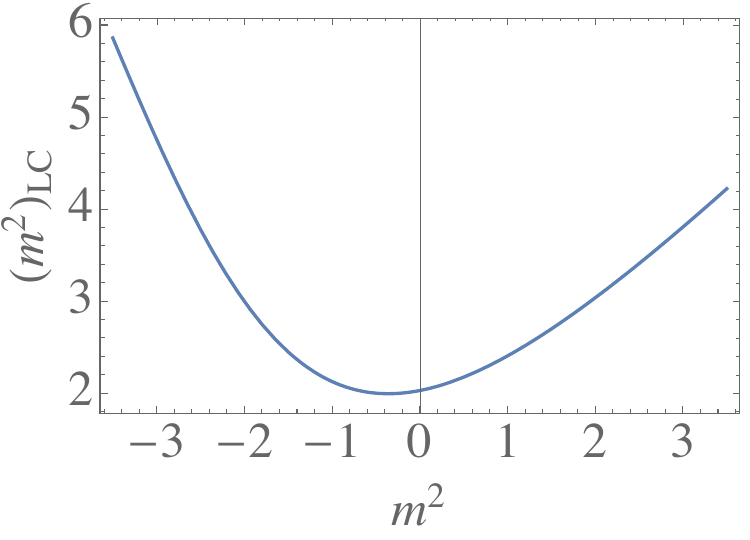}
 \includegraphics[width=0.4\textwidth]{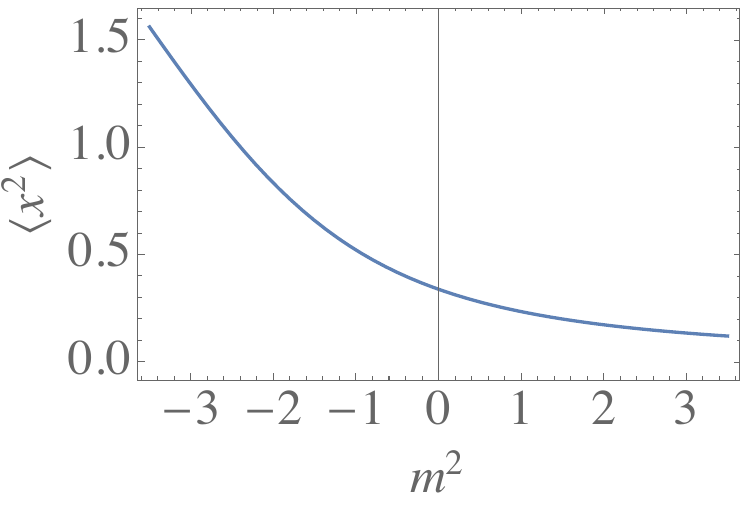}
\caption{{\it Left}: Plot of $m^2_{\rm LC} \equiv m^2 + 6 \< x^2\>$ for the integral (\ref{eq:0dInt}).  The important feature is that $m_{\rm LC}^2$ is not an invertible function of $m^2$.  {\it Right}: Plot of $\<x^2\>$ as a function of $m^2$.  If the perturbative procedure of absorbing plant diagrams into counter-terms were well-defined nonperturbatively, then $\<x^2\>$ would be the same for any two values of $m^2 $ that corresponded to the same $m_{\rm LC}^2$ in the left plot. }
\label{fig:Burkardt0d}
\end{center}
\end{figure}

As before, at large $m^2$ the function is monotonic (and in this case it is monotonic for all positive $m^2$).  However, at negative\footnote{Unlike in higher dimensions, there is no phase transition here at $m^2=0$, and the integral and its moments (like $\<x^2\>$) are well-defined, smooth functions of $m^2$ even across the point $m^2=0$.}  $m^2$, we see that the function turns back up, predicting that the integral should give the same result for multiple values of $m^2$.    From the plot of $\< x^2\>$ in Fig.~\ref{fig:Burkardt0d}, we see that this prediction is false.

Essentially what the ``effective'' integral with plant diagrams subtracted out is doing is defining, as a function of the mass and the counterterm,
\be
Z_{\delta}(m^2_{\rm LC}) \equiv \int_{-\infty}^\infty dx e^{- m_{\rm LC}^2 x^2 - (-\delta x^2 + x^4)}
\ee
such that $\delta$ removes all plant diagrams.  
Manifestly, $Z_\delta$ is related to the original integral by
\be
Z_{\delta}(m^2 + \delta)  = Z(m^2) .
\ee
The perturbation series of $Z_\delta$ differs from that of $Z$, since the former is expanded in inverse powers of $m_{\rm LC}$:
\be
Z_\delta(m^2_{\rm LC} ) = \frac{\sqrt{\pi}}{m_{\rm LC}} \left( 1+ \frac{\delta}{2 m^2_{\rm LC}} + \frac{3 (-2+\delta^2)}{8 m_{\rm LC}^2} + \dots \right)
\ee
Continuing with the analogy, we are interested in choosing $\delta$ to eliminate all of the plant diagrams.  To do this, we take
$
\delta = \delta(m) =  6 \< x^2 \>_{m^2} .
$
In perturbation theory, this is just
\be
\delta(m) = \frac{3}{m^2} \left( 1- \frac{3}{m^4} + \frac{24}{m^8} - \frac{297}{m^{12}}+ \dots \right)
\ee
However, defining $Z_{\delta(m)}(m_{\rm LC})$ at the nonperturbative level requires determining $\delta(m)$ as a function of $m_{\rm LC}$, which in turn requires inverting $m$ as a function of $m_{\rm LC}$.  So while the procedure of defining a $Z_{\delta(m)}(m_{\rm LC})$ with plant diagrams subtracted makes sense in perturbation theory, the point is that it does not make sense nonperturbatively (at a minimum, it must be augmented with a choice of the branch of the solution for $m^2(m^2_{\rm LC})$ -- such an augmentation would be going beyond the perturbative prescription, in analogy with how  Borel-resumming the 2d theory is an augmentation going beyond perturbation theory) which is why the validity of (\ref{eq:Burkardt0d}) is restricted to perturbation theory.

\section{Linear Closing of Gap in $O(N)$ Model}
\label{app:ONModel}

In this appendix we will determine the map between ET and LC parameters in the 3D $O(N)$ model at large-$N$.  This is possible to do because
one can resum perturbation theory at large-$N$.  Parametrizing the large-$N$ theory in the presence of a mass deformation in following way,
\be
\CL = &=& \frac{1}{2} (\partial \phi_i)^2 - \frac{1}{2} \frac{m_{\ET}^2}{\lambda} \sigma -  \frac{1}{2} \sigma (\phi_i)^2 + \frac{\sigma^2}{4\lambda},
\ee
we proceed to calculate the contribution to $H_{\eff}$ which corresponds to $m_{\LC}^2$.  This involves resumming the diagrams of Fig.~\ref{fig:bubblesum}.
\begin{figure}[t!]
\begin{center}
 \includegraphics[width=.8\textwidth]{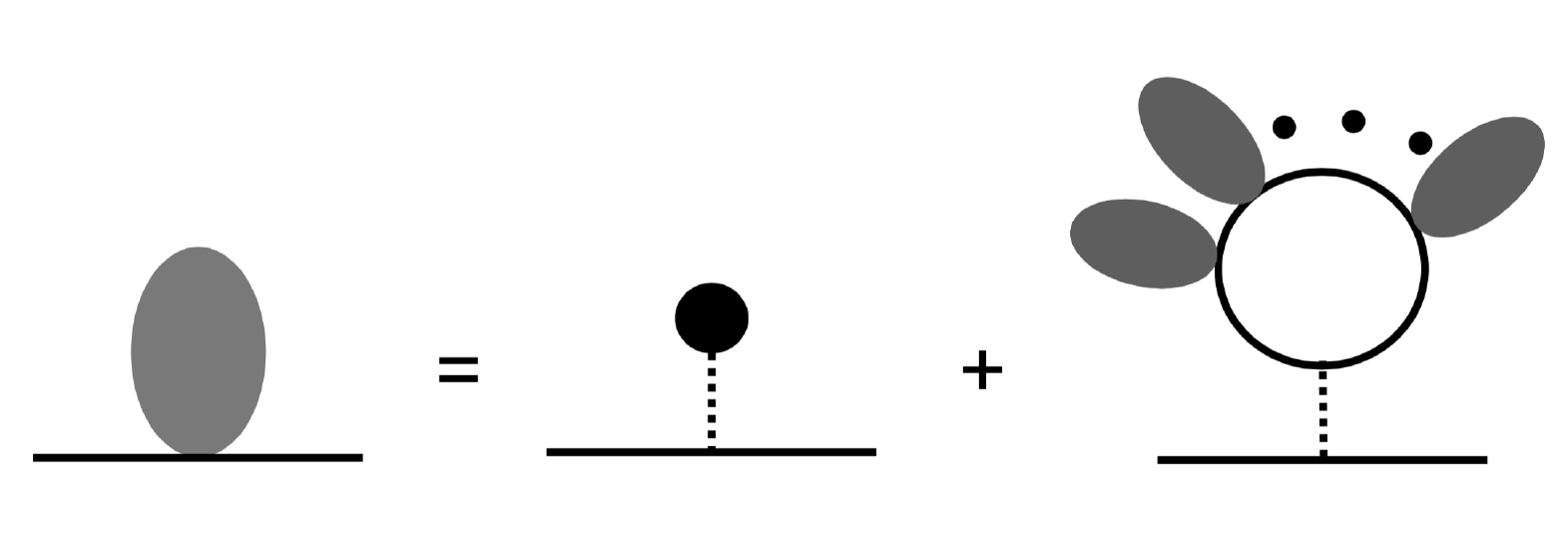}
 \caption{Diagrams contributing to $m_{\LC}^2$. Solid and dashed lines denote $\phi_i$ and $\sigma$ propagators respectively.}
\label{fig:bubblesum}
\end{center}
\end{figure}
The diagrams yield the following standard resummation equation, valid to leading order in $N$ for $\langle\sigma\rangle=m_{\LC}^2$:
\be
m_{\LC}^2 &=& m_{\ET}^2 + \lambda N \int \frac{d^3p}{(2\pi)^3} \left[\frac{1}{p^2 + m_{\LC}^2} -\frac{1}{p^2}\right]
\ee
Here, we have rotated the integral to Euclidean momentum, and have included the effect of normal ordering.  When $m_{\ET}^2 > 0$ and in the limit 
$\lambda \rightarrow \infty$ (i.e.~focusing on the regime near the interacting fixed point) the relation between the parameters is approximately given by
\be
m_{\LC} = \frac{4\pi}{\lambda N}m_{\ET}^2.
\ee 
Let us now recall that in the LC quantization of the $O(N)$ model, the gap closes linearly with $m_{\LC}$.  For instance, the spectral density of the $\phi^2$ operator
along the flow is given by:
\be
\pi\rho_{\phi^2}(q) &=& \frac{\frac{1}{2q}}{\left(1+ \frac{\lambda N}{8 q \pi} \log \left( \frac{q+2m_{\LC}}{q-2m_{\LC}} \right) \right)^2 + \left(\frac{\lambda}{8 q} \right)^2}.
\ee
Thus, as in the case of our 2D scalar example, the LC $O(N)$ model gap closes in a manner which is inconsistent with the power expected from 
the large-$N$ critical exponent of $\nu=1$.  However, again, as in the 2D case, it is the map between ET and LC parameters which resolves this tension.  Indeed, near the critical point the map knows about the $\nu$ exponent:
\be 
\mu_{\gap} \sim m_{\LC} \sim m_{\ET}^2.
\ee

\bibliographystyle{utphys}
\bibliography{BorelBib}

\providecommand{\href}[2]{#2}\begingroup\raggedright\begin{thebibliography}{10}

\bibitem{Weinberg:1966jm}
S.~Weinberg, ``{Dynamics at infinite momentum},''
\href{http://dx.doi.org/10.1103/PhysRev.150.1313}{{\em Phys. Rev.} {\bfseries
  150} (1966) 1313--1318}.

\bibitem{Klauder:1969zz}
H.~Leutwyler, J.~R. Klauder, and L.~Streit, ``{Quantum field theory on
  lightlike slabs},''
\href{http://dx.doi.org/10.1007/BF02826338}{{\em Nuovo Cim.} {\bfseries A66}
  (1970) 536--554}.

\bibitem{Wilson:1994fk}
K.~G. Wilson, T.~S. Walhout, A.~Harindranath, W.-M. Zhang, R.~J. Perry, and
  S.~D. Glazek, ``{Nonperturbative QCD: A weak coupling treatment on the light
  front},'' \href{http://dx.doi.org/10.1103/PhysRevD.49.6720}{{\em Phys. Rev.}
  {\bfseries D49} (1994) 6720--6766},
\href{http://arxiv.org/abs/hep-th/9401153}{{\ttfamily arXiv:hep-th/9401153
  [hep-th]}}.

\bibitem{Brodsky:1997de}
S.~J. Brodsky, H.-C. Pauli, and S.~S. Pinsky, ``{Quantum chromodynamics and
  other field theories on the light cone},''
  \href{http://dx.doi.org/10.1016/S0370-1573(97)00089-6}{{\em Phys. Rept.}
  {\bfseries 301} (1998) 299--486},
\href{http://arxiv.org/abs/hep-ph/9705477}{{\ttfamily arXiv:hep-ph/9705477
  [hep-ph]}}.

\bibitem{Hiller:2016itl}
J.~R. Hiller, ``{Nonperturbative light-front Hamiltonian methods},''
  \href{http://dx.doi.org/10.1016/j.ppnp.2016.06.002}{{\em Prog. Part. Nucl.
  Phys.} {\bfseries 90} (2016) 75--124},
\href{http://arxiv.org/abs/1606.08348}{{\ttfamily arXiv:1606.08348 [hep-ph]}}.

\bibitem{Chang:1968bh}
S.-J. Chang and S.-K. Ma, ``{Feynman rules and quantum electrodynamics at
  infinite momentum},''
\href{http://dx.doi.org/10.1103/PhysRev.180.1506}{{\em Phys. Rev.} {\bfseries
  180} (1969) 1506--1513}.

\bibitem{Yan:1973qg}
T.-M. Yan, ``{Quantum field theories in the infinite momentum frame IV.
  Scattering matrix of vector and Dirac fields and perturbation theory},''
\href{http://dx.doi.org/10.1103/PhysRevD.7.1780}{{\em Phys. Rev.} {\bfseries
  D7} (1973) 1780--1800}.

\bibitem{Maskawa:1975ky}
T.~Maskawa and K.~Yamawaki, ``{The Problem of $P^+ = 0$ Mode in the Null Plane
  Field Theory and Dirac's Method of Quantization},''
\href{http://dx.doi.org/10.1143/PTP.56.270}{{\em Prog. Theor. Phys.} {\bfseries
  56} (1976) 270}.

\bibitem{Tsujimaru:1997jt}
S.~Tsujimaru and K.~Yamawaki, ``{Zero mode and symmetry breaking on the light
  front},'' \href{http://dx.doi.org/10.1103/PhysRevD.57.4942}{{\em Phys. Rev.}
  {\bfseries D57} (1998) 4942--4964},
\href{http://arxiv.org/abs/hep-th/9704171}{{\ttfamily arXiv:hep-th/9704171
  [hep-th]}}.

\bibitem{Yamawaki:1998cy}
K.~Yamawaki, ``{Zero mode problem on the light front},''
\href{http://arxiv.org/abs/hep-th/9802037}{{\ttfamily arXiv:hep-th/9802037
  [hep-th]}}.

\bibitem{Heinzl:2003jy}
T.~Heinzl, ``{Light cone zero modes revisited},''
\href{http://arxiv.org/abs/hep-th/0310165}{{\ttfamily arXiv:hep-th/0310165
  [hep-th]}}.

\bibitem{Beane:2013ksa}
S.~R. Beane, ``{Broken Chiral Symmetry on a Null Plane},''
  \href{http://dx.doi.org/10.1016/j.aop.2013.06.012}{{\em Annals Phys.}
  {\bfseries 337} (2013) 111--142},
\href{http://arxiv.org/abs/1302.1600}{{\ttfamily arXiv:1302.1600 [nucl-th]}}.

\bibitem{Herrmann:2015dqa}
M.~Herrmann and W.~N. Polyzou, ``{Light-front vacuum},''
  \href{http://dx.doi.org/10.1103/PhysRevD.91.085043}{{\em Phys. Rev.}
  {\bfseries D91} no.~8, (2015) 085043},
\href{http://arxiv.org/abs/1502.01230}{{\ttfamily arXiv:1502.01230 [hep-th]}}.

\bibitem{Collins:2018aqt}
J.~Collins, ``{The non-triviality of the vacuum in light-front quantization: An
  elementary treatment},''
\href{http://arxiv.org/abs/1801.03960}{{\ttfamily arXiv:1801.03960 [hep-ph]}}.

\bibitem{Martinovic:2018apr}
L.~Martinovic and A.~Dorokhov, ``{Vacuum loops in light-front field theory},''
\href{http://arxiv.org/abs/1812.02336}{{\ttfamily arXiv:1812.02336 [hep-th]}}.

\bibitem{US}
A.~L. Fitzpatrick, J.~Kaplan, E.~Katz, L.~G. Vitale, and M.~T. Walters,
  ``{Lightcone effective Hamiltonians and RG flows},''
  \href{http://dx.doi.org/10.1007/JHEP08(2018)120}{{\em JHEP} {\bfseries 08}
  (2018) 120},
\href{http://arxiv.org/abs/1803.10793}{{\ttfamily arXiv:1803.10793 [hep-th]}}.

\bibitem{Burkardt}
M.~Burkardt, ``{Light front quantization of the Sine-Gordon model},''
\href{http://dx.doi.org/10.1103/PhysRevD.47.4628}{{\em Phys. Rev.} {\bfseries
  D47} (1993) 4628--4633}.

\bibitem{Burkardt2}
M.~Burkardt, ``{Much ado about nothing: Vacuum and renormalization on the light
  front},''
\href{http://arxiv.org/abs/hep-ph/9709421}{{\ttfamily arXiv:hep-ph/9709421
  [hep-ph]}}.

\bibitem{Serone}
M.~Serone, G.~Spada, and G.~Villadoro, ``{$\lambda \phi^4$ Theory I: The
  Symmetric Phase Beyond NNNNNNNNLO},''
  \href{http://dx.doi.org/10.1007/JHEP08(2018)148}{{\em JHEP} {\bfseries 08}
  (2018) 148},
\href{http://arxiv.org/abs/1805.05882}{{\ttfamily arXiv:1805.05882 [hep-th]}}.

\bibitem{ZinnJustin}
J.~C. Le~Guillou and J.~Zinn-Justin, ``{Critical Exponents from Field
  Theory},''
\href{http://dx.doi.org/10.1103/PhysRevB.21.3976}{{\em Phys. Rev.} {\bfseries
  B21} (1980) 3976--3998}.

\bibitem{Yurov:1989yu}
V.~P. Yurov and A.~B. Zamolodchikov, ``{Truncated conformal space approach to
  scaling Lee-Yang model},''
\href{http://dx.doi.org/10.1142/S0217751X9000218X}{{\em Int. J. Mod. Phys.}
  {\bfseries A5} (1990) 3221--3246}.

\bibitem{Yurov:1991my}
V.~P. Yurov and A.~B. Zamolodchikov, ``{Truncated fermionic space approach to
  the critical 2-D Ising model with magnetic field},''
\href{http://dx.doi.org/10.1142/S0217751X91002161}{{\em Int. J. Mod. Phys.}
  {\bfseries A6} (1991) 4557--4578}.

\bibitem{Coser:2014lla}
A.~Coser, M.~Beria, G.~P. Brandino, R.~M. Konik, and G.~Mussardo, ``{Truncated
  Conformal Space Approach for 2D Landau-Ginzburg Theories},''
  \href{http://dx.doi.org/10.1088/1742-5468/2014/12/P12010}{{\em J. Stat.
  Mech.} {\bfseries 1412} (2014) P12010},
\href{http://arxiv.org/abs/1409.1494}{{\ttfamily arXiv:1409.1494 [hep-th]}}.

\bibitem{Hogervorst:2014rta}
M.~Hogervorst, S.~Rychkov, and B.~C. van Rees, ``{Truncated conformal space
  approach in d dimensions: A cheap alternative to lattice field theory?},''
  \href{http://dx.doi.org/10.1103/PhysRevD.91.025005}{{\em Phys. Rev.}
  {\bfseries D91} (2015) 025005},
\href{http://arxiv.org/abs/1409.1581}{{\ttfamily arXiv:1409.1581 [hep-th]}}.

\bibitem{Rychkov:2014eea}
S.~Rychkov and L.~G. Vitale, ``{Hamiltonian truncation study of the $\phi^4$
  theory in two dimensions},''
  \href{http://dx.doi.org/10.1103/PhysRevD.91.085011}{{\em Phys. Rev.}
  {\bfseries D91} (2015) 085011},
\href{http://arxiv.org/abs/1412.3460}{{\ttfamily arXiv:1412.3460 [hep-th]}}.

\bibitem{Rychkov:2015vap}
S.~Rychkov and L.~G. Vitale, ``{Hamiltonian truncation study of the $\phi^4$
  theory in two dimensions II. The $\mathbb Z_2$-broken phase and the Chang
  duality},'' \href{http://dx.doi.org/10.1103/PhysRevD.93.065014}{{\em Phys.
  Rev.} {\bfseries D93} no.~6, (2016) 065014},
\href{http://arxiv.org/abs/1512.00493}{{\ttfamily arXiv:1512.00493 [hep-th]}}.

\bibitem{Elias-Miro:2015bqk}
J.~Elias-Miro, M.~Montull, and M.~Riembau, ``{The renormalized Hamiltonian
  truncation method in the large $E_T$ expansion},''
  \href{http://dx.doi.org/10.1007/JHEP04(2016)144}{{\em JHEP} {\bfseries 04}
  (2016) 144},
\href{http://arxiv.org/abs/1512.05746}{{\ttfamily arXiv:1512.05746 [hep-th]}}.

\bibitem{Bajnok:2015bgw}
Z.~Bajnok and M.~Lajer, ``{Truncated Hilbert space approach to the 2d
  $\phi^{4}$ theory},'' \href{http://dx.doi.org/10.1007/JHEP10(2016)050}{{\em
  JHEP} {\bfseries 10} (2016) 050},
\href{http://arxiv.org/abs/1512.06901}{{\ttfamily arXiv:1512.06901 [hep-th]}}.

\bibitem{Elias-Miro:2017xxf}
J.~Elias-Miro, S.~Rychkov, and L.~G. Vitale, ``{High-Precision Calculations in
  Strongly Coupled Quantum Field Theory with Next-to-Leading-Order Renormalized
  Hamiltonian Truncation},''
  \href{http://dx.doi.org/10.1007/JHEP10(2017)213}{{\em JHEP} {\bfseries 10}
  (2017) 213},
\href{http://arxiv.org/abs/1706.06121}{{\ttfamily arXiv:1706.06121 [hep-th]}}.

\bibitem{Elias-Miro:2017tup}
J.~Elias-Miro, S.~Rychkov, and L.~G. Vitale, ``{NLO Renormalization in the
  Hamiltonian Truncation},''
  \href{http://dx.doi.org/10.1103/PhysRevD.96.065024}{{\em Phys. Rev.}
  {\bfseries D96} no.~6, (2017) 065024},
\href{http://arxiv.org/abs/1706.09929}{{\ttfamily arXiv:1706.09929 [hep-th]}}.

\bibitem{Hogervorst:2018otc}
M.~Hogervorst, ``{RG flows on $S^d$ and Hamiltonian truncation},''
\href{http://arxiv.org/abs/1811.00528}{{\ttfamily arXiv:1811.00528 [hep-th]}}.

\bibitem{Katz:2013qua}
E.~Katz, G.~Marques~Tavares, and Y.~Xu, ``{Solving 2D QCD with an adjoint
  fermion analytically},''
  \href{http://dx.doi.org/10.1007/JHEP05(2014)143}{{\em JHEP} {\bfseries 05}
  (2014) 143},
\href{http://arxiv.org/abs/1308.4980}{{\ttfamily arXiv:1308.4980 [hep-th]}}.

\bibitem{Katz:2014uoa}
E.~Katz, G.~Marques~Tavares, and Y.~Xu, ``{A solution of 2D QCD at Finite $N$
  using a conformal basis},''
\href{http://arxiv.org/abs/1405.6727}{{\ttfamily arXiv:1405.6727 [hep-th]}}.

\bibitem{Katz:2016hxp}
E.~Katz, Z.~U. Khandker, and M.~T. Walters, ``{A Conformal Truncation Framework
  for Infinite-Volume Dynamics},''
  \href{http://dx.doi.org/10.1007/JHEP07(2016)140}{{\em JHEP} {\bfseries 07}
  (2016) 140},
\href{http://arxiv.org/abs/1604.01766}{{\ttfamily arXiv:1604.01766 [hep-th]}}.

\bibitem{Anand:2017yij}
N.~Anand, V.~X. Genest, E.~Katz, Z.~U. Khandker, and M.~T. Walters, ``{RG flow
  from $\phi^4$ theory to the 2D Ising model},''
  \href{http://dx.doi.org/10.1007/JHEP08(2017)056}{{\em JHEP} {\bfseries 08}
  (2017) 056},
\href{http://arxiv.org/abs/1704.04500}{{\ttfamily arXiv:1704.04500 [hep-th]}}.

\bibitem{Delacretaz:2018xbn}
L.~V. Delacr\'{e}taz, A.~L. Fitzpatrick, E.~Katz, and L.~G. Vitale,
  ``{Conformal Truncation of Chern-Simons Theory at Large $N_f$},''
\href{http://arxiv.org/abs/1811.10612}{{\ttfamily arXiv:1811.10612 [hep-th]}}.

\bibitem{Chabysheva:2018wxr}
S.~S. Chabysheva and J.~R. Hiller, ``{Transitioning from equal-time to
  light-front quantization in $\phi_2^4$ theory},''
\href{http://arxiv.org/abs/1811.01685}{{\ttfamily arXiv:1811.01685 [hep-th]}}.

\bibitem{Bronzin:2018tqz}
S.~Bronzin, B.~De~Palma, and M.~Guagnelli, ``{New Monte Carlo determination of
  the critical coupling in $\phi^4_2$ theory},''
  \href{http://dx.doi.org/10.1103/PhysRevD.99.034508}{{\em Phys. Rev. D}
  {\bfseries 99} no.~3, (2019) 034508},
  \href{http://arxiv.org/abs/1807.03381}{{\ttfamily arXiv:1807.03381
  [hep-lat]}}.

\bibitem{Milsted:2013rxa}
A.~Milsted, J.~Haegeman, and T.~J. Osborne, ``{Matrix product states and
  variational methods applied to critical quantum field theory},''
  \href{http://dx.doi.org/10.1103/PhysRevD.88.085030}{{\em Phys. Rev. D}
  {\bfseries 88} (2013) 085030},
  \href{http://arxiv.org/abs/1302.5582}{{\ttfamily arXiv:1302.5582 [hep-lat]}}.

\bibitem{Kadoh:2018tis}
D.~Kadoh, Y.~Kuramashi, Y.~Nakamura, R.~Sakai, S.~Takeda, and Y.~Yoshimura,
  ``{Tensor network analysis of critical coupling in two dimensional $\phi^{4}$
  theory},'' \href{http://dx.doi.org/10.1007/JHEP05(2019)184}{{\em JHEP}
  {\bfseries 05} (2019) 184}, \href{http://arxiv.org/abs/1811.12376}{{\ttfamily
  arXiv:1811.12376 [hep-lat]}}.

\bibitem{Burkardt:2016ffk}
M.~Burkardt, S.~S. Chabysheva, and J.~R. Hiller, ``{Two-dimensional light-front
  $\phi^4$ theory in a symmetric polynomial basis},''
  \href{http://dx.doi.org/10.1103/PhysRevD.94.065006}{{\em Phys. Rev.}
  {\bfseries D94} no.~6, (2016) 065006},
\href{http://arxiv.org/abs/1607.00026}{{\ttfamily arXiv:1607.00026 [hep-th]}}.

\bibitem{Harindranath:1988zt}
A.~Harindranath and J.~P. Vary, ``{Stability of the Vacuum in Scalar Field
  Models in $1+1$ Dimensions},''
\href{http://dx.doi.org/10.1103/PhysRevD.37.1076}{{\em Phys. Rev.} {\bfseries
  D37} (1988) 1076--1078}.

\bibitem{Romatschke:2019rjk}
P.~Romatschke, ``{Simple non-perturbative resummation schemes beyond
  mean-field: case study for scalar $\phi^4$ theory in 1+1 dimensions},''
  \href{http://dx.doi.org/10.1007/JHEP03(2019)149}{{\em JHEP} {\bfseries 03}
  (2019) 149}, \href{http://arxiv.org/abs/1901.05483}{{\ttfamily
  arXiv:1901.05483 [hep-th]}}.

\bibitem{Elliott:2014fsa}
B.~Elliott, S.~S. Chabysheva, and J.~R. Hiller, ``{Application of the
  light-front coupled-cluster method to $\phi^4$ theory in two dimensions},''
  \href{http://dx.doi.org/10.1103/PhysRevD.90.056003}{{\em Phys. Rev.}
  {\bfseries D90} no.~5, (2014) 056003},
\href{http://arxiv.org/abs/1407.7139}{{\ttfamily arXiv:1407.7139 [hep-ph]}}.

\bibitem{Chabysheva:2016ehd}
S.~S. Chabysheva and J.~R. Hiller, ``{Light-front $\phi_2^4$ theory with
  sector-dependent mass},''
  \href{http://dx.doi.org/10.1103/PhysRevD.95.096016}{{\em Phys. Rev.}
  {\bfseries D95} no.~9, (2017) 096016},
\href{http://arxiv.org/abs/1612.09331}{{\ttfamily arXiv:1612.09331 [hep-th]}}.

\bibitem{Bosetti:2015lsa}
P.~Bosetti, B.~De~Palma, and M.~Guagnelli, ``{Monte Carlo determination of the
  critical coupling in $\phi^4_2$ theory},''
  \href{http://dx.doi.org/10.1103/PhysRevD.92.034509}{{\em Phys. Rev. D}
  {\bfseries 92} no.~3, (2015) 034509},
  \href{http://arxiv.org/abs/1506.08587}{{\ttfamily arXiv:1506.08587
  [hep-lat]}}.

\bibitem{Schaich:2009jk}
D.~Schaich and W.~Loinaz, ``{An improved lattice measurement of the critical
  coupling in $\phi_2^4$ theory},''
  \href{http://dx.doi.org/10.1103/PhysRevD.79.056008}{{\em Phys. Rev. D}
  {\bfseries 79} (2009) 056008},
  \href{http://arxiv.org/abs/0902.0045}{{\ttfamily arXiv:0902.0045 [hep-lat]}}.

\bibitem{Sugihara:2004qr}
T.~Sugihara, ``{Density matrix renormalization group in a two-dimensional
  $\lambda \phi^4$ Hamiltonian lattice model},''
  \href{http://dx.doi.org/10.1088/1126-6708/2004/05/007}{{\em JHEP} {\bfseries
  05} (2004) 007}, \href{http://arxiv.org/abs/hep-lat/0403008}{{\ttfamily
  arXiv:hep-lat/0403008}}.

\end{thebibliography}\endgroup

\end{document}